%% file: main.tex
\documentclass[11pt]{tcibook}
\usepackage{fancyhea}
\usepackage{work}
\usepackage{bm}       
\usepackage{graphicx}
\usepackage{multirow}
\usepackage{lineno}
\usepackage{threeparttable}
\usepackage[normalem]{ulem}
\usepackage{color}
\usepackage{cite}
\usepackage[colorlinks = true,
            linkcolor = blue,
            urlcolor  = blue,
            citecolor = blue,
            anchorcolor = blue]{hyperref}

\usepackage{amssymb}
\input workshopsymbols.tex     

\usepackage{moresize}
\usepackage{titling}

\title{Snowmass Theory Frontier Report}
\usepackage{authblk}

\input{authors.tex}
\date{}
\begin{document}

\parindent=0pt
\parskip=6pt
\setlength{\evensidemargin}{0pt}
\setlength{\oddsidemargin}{0pt}
\setlength{\marginparsep}{0.0in}
\setlength{\marginparwidth}{0.0in}
\marginparpush=0pt

\maketitle

\pagenumbering{roman}

\setcounter{chapter}{6} 

\tableofcontents

\clearpage
\pagenumbering{arabic}

\section*{Executive Summary}

Theory is essential to the field of particle physics, unifying the frontiers and underlying our understanding of the most basic building blocks of Nature. Theory produces transformative science both in connection to {\it Projects} and in its {\it own right}. Below we summarize the various indispensible roles theory plays within particle physics. 
\vspace{-6pt} 
\begin{itemize} \itemsep2pt\parskip2pt \parsep2pt 
\item Theory  is central to the motivation, analysis, and interpretation of experiments. 
\item Theory lays the foundations for future experiments. 
\item Theory facilitates our explorations of Nature in regimes that experiments cannot (yet) reach. 
\item Theory connects particle physics to gravity, cosmology, astrophysics, nuclear physics, condensed matter, atomic, molecular, and optical (AMO) physics, and mathematics. 
\item Theory extends the boundaries of our understanding of Nature, incorporates new perspectives (such as quantum information), and develops new techniques relevant to experiment. 
\item Theorists are responsive, pinpointing novel directions based on recent experimental data, proposing future experiments, and developing the precision tools necessary to interpret them, all with the aim of maximizing experimental impact. 
\item Theorists play a leading role in disseminating new ideas to a wider public and serve as role models for aspiring young scientists.
\end{itemize}

  Together, {\it fundamental theory, phenomenology} and {\it computational theory} form a vibrant, balanced and interconnected scientific ecosystem closely aligned with experiment. The past decade has witnessed significant advances across the many facets of the field, with immense promise in the years to come. Here we list some of the highlights. 

{\bf Fundamental Theory}:
\vspace{-6pt}
\begin{itemize}\itemsep2pt\parskip2pt \parsep2pt 
\item New perturbative and non-perturbative techniques (ranging from the double copy structure of scattering amplitudes to the advent of diverse bootstrap methods) have vastly expanded our knowledge of quantum field theory, in tandem with complementary advances in lattice field theory quantifying non-perturbative properties in theories of interest. 
\item A deeper understanding of holography and insights from quantum information have breathed new life into the longstanding quest for a complete theory of quantum gravity. 
\item New effective field theories have facilitated applications of quantum field theory (QFT) to high-energy scattering, Higgs physics, large-scale structure, inflation, dark matter detection, and gravitational waves. 
\end{itemize}
{\bf Phenomenology}:
\vspace{-6pt}
\begin{itemize}\itemsep2pt\parskip2pt \parsep2pt
\item The search for physics beyond the Standard Model has broadened considerably with the advent of novel concepts like neutral naturalness or cosmological selection of the electroweak vacuum. 
\item Dark matter theory is undergoing a renaissance with the exploration of the full range of allowed dark matter masses, numerous portals to dark sectors, and novel interaction mechanisms. Advances in dark matter phenomenology have gone hand in hand with new proposals for dark matter experiments, often envisioned and implemented by theorists. 
\item The discovery of gravitational waves has catalyzed rapid progress in precision calculations via scattering amplitudes and inspired the use of gravitational waves to study particle physics inaccessible via planned colliders. 
\item
In precision collider theory the calculation of cross sections to the next-to-next-to-leading order (NNLO) and beyond in QCD has now become possible, unlocking the door to unprecedented tests of the SM. This topic is a prime example where cross-fertilization between phenomenology and fundamental theory has accelerated progress in both areas. 
\item 
Precision flavor physics has been the cradle of many of our most powerful effective field theory tools including heavy quark effective theory and soft-collinear effective theory. Taking advantage of the large amount of experimental data, they are now enabling advanced theoretical analyses to obtain constraints on promising beyond the Standard Model (BSM) candidate theories, maximizing the discovery potential of the experiments. 
\item An explosion of theoretical activity in collider phenomenology has led to many new collider observables including many forms of jet substructure and the emerging field of multi-point correlators, employing widespread innovations in computational theory to leverage machine learning and artificial intelligence. 
\item A theory-driven, coordinated program combining nuclear effective theory, lattice QCD, perturbative QCD, and event generation to quantify the multi-scale nuclear cross sections at the needed precision level has been launched, that will allow us to unlock the full potential of the present and future neutrino physics program. 
\end{itemize}
{\bf Computational Theory}: 
\vspace{-6pt}
\begin{itemize}\itemsep2pt\parskip2pt \parsep2pt
\item Lattice QCD has become a powerful tool for precision physics, yielding precise SM predictions that reveal surprising new tensions in quark- and lepton-flavor physics. 
The scope of lattice QCD is undergoing a rapid expansion, promising to provide quantitative access in the coming decade to important new observables, including those involving nucleons and nuclei. 

\item Innovations in the theory of machine learning (AI/ML), such as the development of (gauge) symmetry equivariant architechtures, driven by theoretical work lattice field theory, collider phenomenology, and other areas, may have transformative impact on computations in high energy theory and beyond. 

\item Recent dedicated efforts to develop the methods and theoretical foundations for quantum simulations of quantum field theories relevant to high energy theory are already yielding intriguing results on currently available hardware, offering great promise for computations of classically intractable problems in the decades to come. 
\end{itemize}
The US theory community has played a leading role in all of these endeavors, and will sustain its position of international leadership with continued support. To this end,
\vspace{-6pt}
\begin{itemize}\itemsep2pt\parskip1pt \parsep1pt
    \item The United States and its partners should emphatically support a broad and balanced program of theoretical research covering the entirety of high-energy physics, from fundamental to phenomenological to computational topics, both in connection to experiment and in its own right.
    \item The theory community is most effective as part of a more balanced HEP program of Projects and Research, as both are essential to the health of the field.  Within this balance, support for people is vitally important, as they constitute the primary infrastructure of Research endeavors.
    \item The effectiveness of the theory community as a part of the broader HEP enterprise is enhanced by targeted bridge-building initiatives that connect theory to experiment, such as the Neutrino Theory Network.  
    Support for ongoing and emerging initiatives will strengthen connections to experiment and sharpen focus on key objectives.
    \item It is vital to maintain a program in theory (and across HEP) that trains students and junior scientists, providing them with continuing physics opportunities that empower them to contribute to science.
    \item The success of our endeavors depends on ``$4 \pi$'' coverage in identifying and cultivating talent at all career stages. This requires strengthening our commitment to improving diversity, equity, inclusion, and accessibility in the field.
\end{itemize}

\clearpage
\section{Introduction}

The profound value of theoretical physics is perhaps most apparent in the twin scientific revolutions of the 20th century: relativity and quantum mechanics. At the time of their formulation, neither was thought to have any practical consequences; a century later, they form the foundation of modern technology. Theory lays the groundwork for experiments and technologies of the future. But its value should not be measured exclusively by its potential for future application. Theory addresses our essential human curiosity to understand nature at the deepest level, encompassing not only the actual laws of nature but also the conceivable ones. 

The history of the discovery of the Higgs boson and the current and future precision measurements of its properties serve as a perfect illustration of the multiple roles of theory in the particle physics ecosystem. In the early 1960s, the concept of the Higgs mechanism and a Higgs boson was purely theoretical -- it was proposed to reconcile a theoretical paradox: how can electroweak symmetry be a basic underlying organizing principle, yet also be broken at the same time. At the time of its invention it seemed impossible to  actually detect the Higgs boson; it was fundamental theory making predictions about Nature in regions we could not yet probe.  When it became clear in the early 1970s that the Higgs mechanism was indeed likely to underlie the electroweak interactions, theorists started to seriously explore all possible ways of discovering it experimentally, culminating in {\it The Higgs Hunter's Guide} -- an entire book devoted to the various approaches to Higgs detection depending on its detailed properties. 
These theoretical efforts have motivated and driven a three-decades-long experimental program, first at LEP, then at the Tevatron and the LHC, 
that famously led to the Higgs discovery in 2012. Theoretical input has strongly influenced the experimental program, even at the level of detector design: the electromagnetic calorimeters in the LHC experiments were famously designed such that a Higgs decaying to two photons as predicted by theory could be observed and used as a discovery channel.   The quest to understand the nature of the Higgs boson by determining its properties at the LHC and at future experiments  is also strongly intertwined with cutting-edge theory research. 
Recent advances in precision collider physics have made calculations of the production cross sections at the needed N$^3$LO order in perturbative QCD possible. Likewise, the partial decay widths are known through sufficiently high loop order, enabling precise theoretical predictions, when coupled with inputs for the quark masses and strong coupling ($\alpha_s$) determined from lattice QCD. 
In this respect, theory is providing key tools necessary to analyze the results of experiments. This one story nicely illustrates all the major roles of theory -- from fundamental theoretical concept, to motivating and laying the groundwork for future experiments, to helping analyze and interpret the actual experimental data.  

This report summarizes the recent progress and promising future directions in theoretical high-energy physics (HEP) identified within the Theory Frontier of the 2021 Snowmass Process. The goal of the Theory Frontier is to articulate the recent advances and future opportunities in all aspects of theory relevant to HEP, including particle theory, fundamental theory, cosmological and astro-particle theory, and quantum information science. To this end, the Theory Frontier is comprised of eleven interwoven topical groups spanning string theory, quantum gravity, and black holes (TF01 \cite{TF01}); effective field theory techniques (TF02 \cite{TF02}); conformal field theory and formal quantum field theory (TF03 \cite{TF03}); scattering amplitutes (TF04 \cite{TF04}); lattice gauge theory (TF05 \cite{TF05}); theory techniques for precision physics (TF06 \cite{TF06}); collider phenomenology (TF07 \cite{TF07}); beyond-the-Standard Model model building (TF08 \cite{TF08}); astro-particle physics and cosmology (TF09 \cite{TF09}); quantum information science (TF10 \cite{TF10}); and the theory of neutrino physics (TF11 \cite{TF11}). The activities of the Theory Frontier complement theory as it appears in the other Snowmass frontiers, and the essential role of theory is further articulated in those frontier summaries. 

The central thesis of this report is that theory is a vital and vibrant part of HEP, producing transformative science both in connection to Projects and in its own right. It is essential to the motivation, analysis, and interpretation of experiments; lays the foundations for future experiments; drives of the development and implementation of enabling technologies; and advances our understanding of Nature in regimes that experiments cannot (yet) reach. The theory community in the United States is world-leading, a position that cannot be taken for granted but may be maintained with sustained support and stewardship.

\begin{figure}[htbp] 
   \centering
   \includegraphics[width=0.45\textwidth]{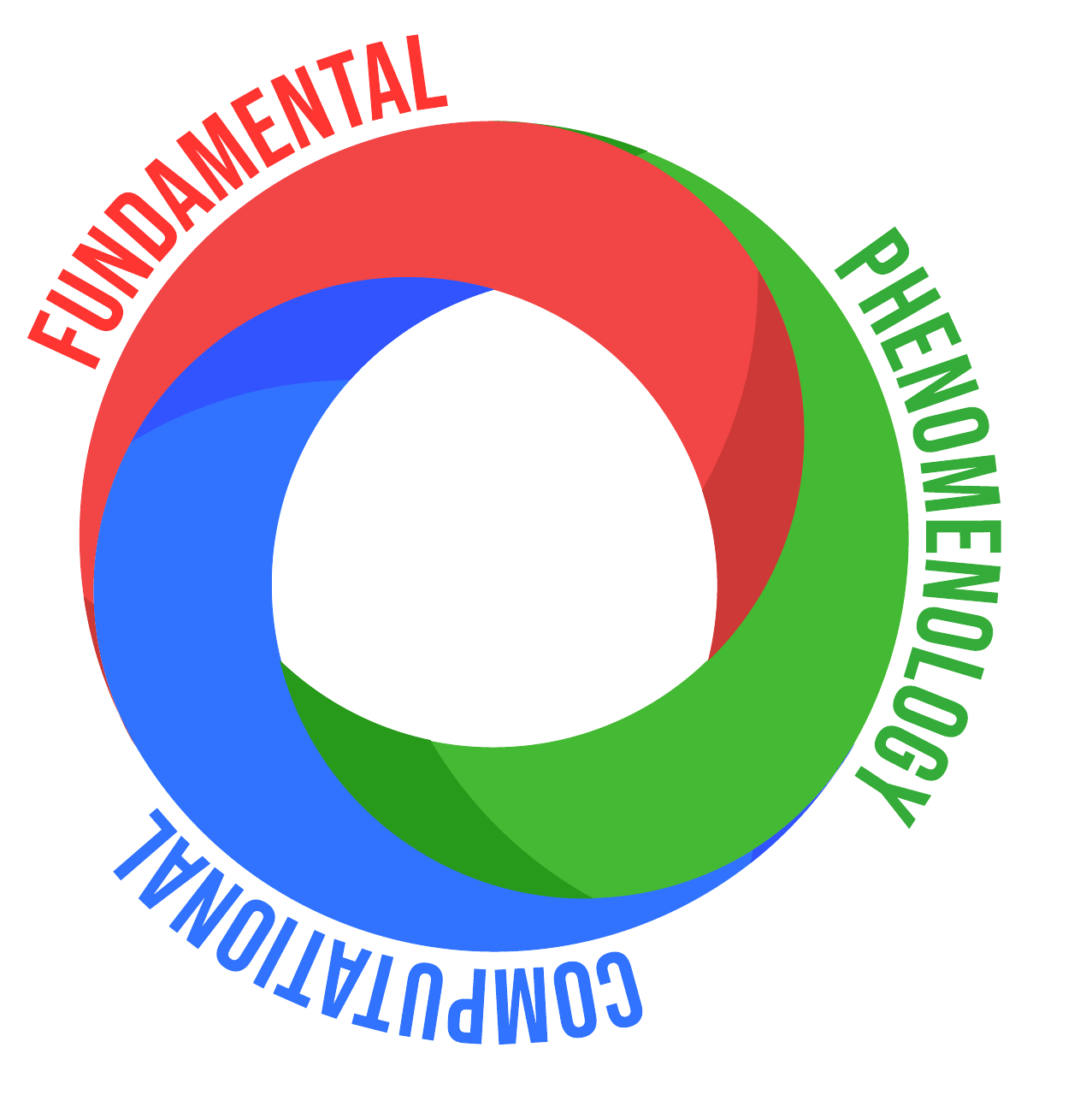} 
   \caption{Fundamental theory, phenomenology, and computational theory form a vibrant, balanced, and interconnected scientific ecosystem (represented here as an umbilic torus, whose three-sided cross sections belie its single face) spanned by the eleven topical groups of the Theory Frontier. Image credit: P.~Tanedo.}
   \label{fig:theoryfig}
\end{figure}

The report itself is organized around three interconnected themes: fundamental theory, phenomenology, and computational theory. The interwoven nature of these themes is illustrated schematically in Fig.~\ref{fig:theoryfig}.
Fundamental theory is often also referred to as formal theory. It seeks deep understanding of the theoretical and mathematical structures that underlie our modern description of Nature and includes directions which are not (or not yet) directly connected to experimentally testable consequences. Phenomenology explores the connections between theoretical principles and experimental observations to gain deeper insights into the theories realized in Nature. Computational theory seeks to quantitatively test our theoretical descriptions of physical phenomena and gain new insights into fundamental aspects of the underlying theories, through developing and deploying ever more powerful computational methods. It bears emphasizing that these themes are deeply interconnected and effectively inseparable, with developments in each theme strongly influencing the others.
While the impact of fundamental theory on phenomenology and computational theory is perhaps the most apparent, developments such as the modern amplitudes program, the revival of the conformal bootstrap, and the development of exact techniques in supersymmetric gauge theories speak to the influence of phenomenology on fundamental theory. Similarly, computational theory has considerably influenced phenomenology by revealing key non-perturbative properties of the Standard Model and its extensions, while new computational tools have enabled rapid progress in diverse aspects of fundamental theory. 

Recent developments and promising future directions in each of these three themes have been synthesized from the summary documents produced by each of the eleven topical groups within the Theory Frontier \cite{TF01,TF02,TF03,TF04,TF05,TF06,TF07,TF08,TF09,TF10,TF11}. Given the tightly interwoven nature of progress across the field, each theme combines developments from multiple topical groups, and developments in any one topical group are often reflected in more than one theme. Loosely speaking, the section on fundamental theory primarily covers highlights from TF01, TF02, TF03, TF04, and TF10; the section on phenomenology primarily covers highlights from TF02, TF04, TF06, TF07, TF08, TF09, TF10 and TF11; and the section on computational theory primarily covers highlights from TF05 and TF10 and includes computational developments in other topical groups. This report, and the topical group reports on which it is based, consolidate community input from 150 contributed papers and a dedicated Theory Frontier conference, and reflect refinement by several rounds of community feedback.

\section{Fundamental theory} \label{sec:funth}

It is the great privilege of 21st century physicists to explore the frameworks that arise at the intersection of relativity and quantum mechanics: quantum field theory and quantum gravity. Although textbook methods of quantum field theory have long lent themselves to perturbative expansions at low orders, recent progress has led to new formulations of quantum field theory, new connections between seemingly disparate theories, and the exploration of heretofore-inaccessible regimes. Meanwhile, the decades-long search for a theory that fully combines general relativity and quantum mechanics has been animated by new insights from string theory and quantum information theory. Progress in quantum field theory and quantum gravity has been tightly intertwined by the revelation of holography (the idea that quantum gravitational physics in a region of space is encoded by a quantum field theory living on the boundary of this region) and the surprising double copy relation between scattering amplitudes in gravity and gauge theory. Progress in both areas has contributed to new understanding across wide swathes of physics and mathematics.

\subsection{Quantum field theory} \label{sec:qft}

The historical approach to the union of special relativity and quantum mechanics gave rise to the wildly successful formulation of quantum field theory in terms of path integrals formulated with an explicit measure and Lagrangian. This perspective lends itself to perturbative expansions that underlie much of our current understanding of the Standard Model, but becomes increasingly cumbersome at high orders in perturbation theory and fails entirely in non-perturbative regimes. Both limitations must be overcome in order to completely describe the Standard Model in general, and QCD in particular. Rigorously establishing the existence of Yang-Mills theory and the phenomenon of confinement remains an open question of profound significance. More broadly, the existence of non-Lagrangian QFTs suggests that the historical formulation is essentially incomplete and the space of possible theories remains to be fully explored.

In the past decade, extraordinary progress has been made in developing new non-perturbative approaches to quantum field theory and extending the reach of perturbative ones. Quantum information provides a new perspective on quantum field theory, explored further in Sec.~\ref{sec:qis}. Symmetries have also played a central role in this progress, including not only familiar spacetime symmetries (such as Lorentz symmetry, conformal symmetry, or superconformal symmetry) and conventional group-based global symmetries, but also more recent generalizations including higher form symmetries, higher group symmetries, subsystem symmetries, and non-invertible symmetries. Although such generalized symmetries have only been discovered quite recently, they appear to be abundantly realized in nature, with considerable relevance to both high energy and condensed matter physics (and even the familiar Standard Model). The implications and applications of generalized symmetries are likely to drive considerable progress in the coming years. 

The quantum field theories respecting a given symmetry comprise a deeply interconnected ``theory space'' populated by both familiar theories and heretofore-unknown ones.  The ultimate goals of the modern QFT research program include the characterization of all possible symmetries governing QFTs, the comprehensive mapping of the associated theory spaces, and the construction of their physical observables. Attaining these goals will profoundly expand our understanding of field theories appearing in nature and shed new light on the microscopic physics underlying the Standard Model.

\subsubsection{Non-perturbative developments} 

The constraints imposed by symmetries are powerful enough to allow modern ``bootstrap'' methods to construct observables non-perturbatively and proscribe the space of possible theories. The bootstrap program has been particularly successful in the context of conformal field theories (CFTs), i.e.~QFTs with additional symmetry under the conformal group. Such CFTs are ubiquitous, emerging as the long-distance description of many systems in both condensed matter and high energy physics, forming the starting point for the formulation of many QFTs, and featuring prominently in holographic descriptions of quantum gravity via the AdS/CFT correspondence. Recent progress in the conformal boostrap program has been driven by both numerical and analytical advances. On the numerical side, the constraints imposed by conformal symmetry have been reformulated in terms of convex optimization problems. This has allowed world-record computations of critical exponents in physically relevant CFTs such as the 3d Ising model and provided a systematic approach to proscribing the space of consistent CFTs. On the analytic side, diverse Lorentzian methods have led to new solutions of CFTs in various limits, as well as new bounds on event shapes, energy distributions, and S-matrices. Progress in the conformal bootstrap is further synergistic with a growing number of lattice studies of the conformal transition and conformal window in both supersymmetric and non-supersymmetric Yang-Mills theory. Ongoing efforts to merge numerical and analytic approaches are bound to yield considerable progress in the years to come.

Related methods underlie the recent renaissance of the S-matrix bootstrap. Here the imposition of bedrock principles such as analyticity, unitarity, and Lorentz invariance has allowed the space of consistent S-matrices to be mapped out {\it nonperturbatively} using a variety of optimization methods. Validation of this approach in 2d theories has sparked ongoing efforts in higher-dimensional theories, giving rise to impressive constraints on both pion scattering in 4d and graviton scattering in 10d. These results dovetail with newly-appreciated positivity bounds in effective field theories and with the explosion of perturbative techniques for the computation of S-matrices within the amplitudes program. The recent extension of bootstrap methods to correlation functions in de Sitter space has led to the emergence of a cosmological bootstrap program with far-reaching implications for inflation and early-universe cosmology.   

Points in theory space are often related as a function of scale, a generalization of the familiar Wilsonian notion of renormalization group (RG) flow between fixed points. Such flows often proceed from CFTs in the ultraviolet to gapped phases in the infrared, raising the question of whether the physical properties of the IR gapped phase can be reconstructed from the data of the UV CFT and the deforming perturbation. In principle this amounts to diagonalizing the Hamiltonian of the perturbed theory, implemented numerically in the revived framework of Hamiltonian Truncation. Hamiltonian Truncation and related methods have been successfully applied to a range of 2d QFTs, with extensions to 3d and 4d theories underway. Ultimately, this may yield a new nonperturbative scheme for computing the spectrum of 4d QCD starting from a (potentially solvable) UV CFT.  

The mapping of supersymmetric theory spaces is especially advanced, particularly for theory spaces governed by (extended) superconformal symmetry. Extensive progress has been made in the classification of both $\mathcal{N} = (2,0)$ and $\mathcal{N}=(1,0)$ superconformal field theories (SCFTs) in six dimensions, as well as SCFTs in three, four, and five dimensions obtained by compactification. There are many interesting SCFTs with lesser amounts of supersymmetry in diverse dimensions, related by a rich web of dualities. While the full classification of these theories is out of reach at present, it provides a motivated target for the rapidly-developing bootstrap approaches discussed above. Deformations of these theories are bound to lend new insight into non-supersymmetric theories such as QCD.

\subsubsection{Perturbative developments and the amplitudes program} 

Scattering processes are key experimental and theoretical probes of quantum field theories, underlying much of our modern understanding of elementary particles. Scattering amplitudes span all aspects of high-energy physics, ranging from collider physics to mathematical physics to string theory and supergravity, and even recently extending into the realm of gravitational-wave physics. The study of scattering amplitudes (typically, but not exclusively, in perturbation theory) has offered remarkable insight into the structure of quantum field theories and enabled the high-precision theoretical predictions necessary to interpret modern experiments. Recent progress in scattering amplitudes has proceeded through a sort of ``virtuous cycle'' in which calculating quantities of experimental or theoretical interest fuels new discoveries about the properties of quantum field theories. These discoveries, in turn, enable the development of new computational techniques to generate additional insights. 

Scattering amplitudes possess a deep structure that is obscured by traditional Feynman diagrammatic methods. In recent years, considerable progress has been made in revealing this structure, including the description of amplitudes in terms of algebraic curves in twistor space; the development of unitarity-based iterative methods for constructing loop-level amplitudes from tree-level ones; the recursive construction of tree-level amplitudes; the discovery of a hidden ``dual conformal'' symmetry in planar maximally supersymmetric Yang-Mills theory; the representation of amplitudes in geometric terms (such as the Amplituhedron); and the revelation that amplitudes exhibit a duality between color and kinematics. These novel descriptions have, in turn, led to greatly improved computational methods. 

In the context of fundamental theory, the efficient calculation of scattering amplitudes provides a form of theoretical data that tests new ideas and reveals surprising connections. This has driven significant advances in our understanding of both quantum field theory and quantum gravity. Nowhere is this more apparent than in the study of maximally supersymmetric Yang-Mills (SYM) theory in the planar limit. In some sense the simplest quantum field theory, it provides an ideal testing ground for amplitudes methods and a toy model for theories such as QCD. The perturbative expansion of maximally supersymmetric Yang-Mills is convergent, with ultraviolet-finite scattering amplitudes exhibiting a hidden infinite-dimensional Yangian symmetry. In recent years, this has allowed extraordinary progress in studying its scattering amplitudes at both weak and strong coupling. The virtuous cycle between new methods, new amplitude calculations, and new structures in quantum field theory is bound to yield further advances in the years to come.

\subsubsection{Effective field theory}

Unsurprisingly, the power and generality of quantum field theory leads to considerable mathematical challenges in its application to the real world. Among other sources of complexity, field theories occurring in Nature often feature hierarchically distinct length scales. But locality allows us to disentangle these scales in order to usefully reduce the computational complexity required for a given prediction. In practice, this amounts to finding an appropriate description for the important physics at a given length scale, with the number of free parameters controlled by symmetries and their importance controlled by some form of power counting. This approach is known as effective field theory (EFT). 

EFTs are indispensable to the application of quantum field theory in the real world, and many quantum field theories realized by Nature are only understood through the lens of EFT. The modern philosophy of EFT originated in HEP with the study of pion physics, but has since become ubiquitous across many fields of physics. Within HEP, EFT has found use in a dizzying array of contexts ranging from dark matter detection to large-scale structure to high-energy scattering to gravitational waves. EFT reasoning also gives rise to many of the ``big questions'' surrounding the Standard Model, such as the naturalness puzzles of the electroweak hierarchy problem, the strong CP problem, and the cosmological constant problem. Moreover, the study of EFT itself has yielded new insights into the constraints arising from bedrock principles such as unitarity and causality. Within the past decade, there has been considerable progress in the systematics of effective field theory, as new techniques (augmented by on-shell methods developed in the amplitudes program) have facilitated the explicit construction of EFTs and revealed their underlying simplicity. Growing connections between effective field theory and the amplitudes program are likely to yield further significant advances in the years to come.

Dark matter (DM) detection leverages EFT for both direct and indirect measurements, including systematizing low-energy DM-nuclei interactions and re-summing large logarithms in DM annihilation. In cosmology, EFTs have been employed with great success to study Large Scale Structure (LSS) and inflation. Among other results, EFT calculations have produced state-of-the-art predictions for cosmological parameters extracted from LSS data that are competitive with those from the cosmic microwave background. In models of inflation, EFT techniques have been used to expand analysis strategies and place novel bounds on non-gaussianities.

The role of EFT in collider physics and phenomenology continues to grow. Soft-Collinear Effective Theory (SCET), responsible for greatly improving our understanding of factorization in high-energy scattering, remains central to precision calculations at colliders. The discovery of the Higgs boson at the LHC highlighted the importance of treating the Standard Model as an effective field theory, leading to rapid developments in the Standard Model EFT (SMEFT) and Higgs EFT (HEFT). These EFTs have transformed the experimental interpretation of Higgs coupling measurements and fuelled considerable exploration (aided in part by the amplitudes program) of their novel field-theoretic properties.

The expansion of EFT techniques into adjacent fields has led to new breakthroughs and produced fruitful feedback back to HEP. For example, EFT approaches in condensed matter physics have shed light on equilibrium phenomena, hydrodynamics, and (non) Fermi liquids, while EFT techniques underlie recent inter-disciplinary study of exotic states of matter known as fractons. A new EFT plays a key role in the burgeoning field of gravitational wave astronomy by organizing the signatures of binary inspirals, with systematic improvements from the amplitudes program.

Beyond its many applications, the framework of effective field theory has been considerably refined in recent years by an improved understanding of the constraints imposed by unitarity and causality, driven in part by developments in the amplitudes program. While the historical approach to EFT (including all effective operators involving the light degrees of freedom, consistent with the symmetries, with arbitrary Wilson coefficients) has been used to great effect, there is a growing appreciation that the apparently-vast space of operator coefficients is sharply proscribed by the existence of a consistent UV completion. Imposing unitarity and causality leads to constraints on the derivative expansions of scattering of amplitudes, often in the form of positivity bounds on Wilson coefficients. This both deepens our understanding of the systematics of EFT and provides powerful guidance to experiments whose results can be framed in terms of EFTs. Rapid progress has been made in the space of the past few years, while the complete exploration of EFT constraints arising from consistent UV completion is a promising target for the years to come.

\subsection{Quantum gravity} \label{sec:grav}

The union of general relativity and quantum mechanics remains one of the main outstanding challenges of theoretical physics. The semiclassical quantization of general relativity is viable as an effective field theory, but breaks down in situations where the spacetime curvature is not small in Planck units. This includes many singularities of profound interest, such as the initial singularity of the universe and singularities in the interior of black holes.  

Yet the greatest challenges have a tendency to catalyze the greatest progress. Although radio telescope arrays and gravitational wave observatories have taught us a great deal about the existence and environs of black holes in recent years, at present only a theory of quantum gravity can extend across a black hole's event horizon to characterize the singularity at its center. Thought experiments involving the quantum mechanics of black holes have had transformative impacts on our understanding of quantum gravity and the nature of spacetime. 

Semiclassical calculations strongly suggest that a black hole possesses an entropy proportional to its area and evaporates on a timescale proportional to the cube of its mass. These properties lead directly to Hawking's famous information paradox: information about how the black hole was made lies deep in its interior, space-like separated (in the semiclassical picture) from the near-horizon region where evaporation takes place, implying that evaporation destroys this information in contradiction with the unitarity of quantum mechanics. It is by now clear that unitarity cannot be restored by ``small corrections'' to Hawking's original calculation, requiring us to accept the failure of either quantum mechanics or the semiclassical picture of spacetime. The latter implies that spacetime near black holes -- and, by extension, everywhere -- must be emergent. Rendering this precise is a central focus of modern research in quantum gravity. 

The emergence of spacetime can be made concrete in string theory, where the AdS/CFT correspondence has shed light on the emergence of bulk gravitational spacetime from the field theory description on the boundary. A central breakthrough was the Ryu-Takayanagi proposal, later refined into the quantum extremal surface formula, for computing the von Neumann entropy of a boundary state in terms of bulk quantities. This, in turn, revealed that boundary entanglement is a key element in the emergence of bulk spacetime, encoded most precisely in the revelation that the AdS/CFT correspondence can be understood in this context as a quantum error-correcting code. Most recently, these insights led to the discovery that the ``Page curve'' -- the rise and fall of entanglement entropy entailed by the unitary evolution of an evaporating black hole -- can be reliably computed using the quantum extremal surface formula.

Parallel developments have revealed a deep connection between the dynamics of black holes and quantum chaos, namely that simple gravitational calculations serve as probes of chaos in the boundary description. This and related ideas have been explicitly explored in the 2d Sachdev-Ye-Kitaev (SYK) model, thought to be holographically dual to Jackiw-Teitelboim gravity, enabling tests of the chaotic nature of black holes in a solvable regime. Related developments have uncovered connections between computational complexity and the emergence of spacetime, leading to a range of novel insights and ongoing progress. More broadly, the study of quantum chaos in black holes led to a new diagnostic, the out-of-time-order correlator. This can be applied to study chaos in any many-body quantum system, leading to insights ranging from the identification of a characteristic ``butterfly velocity'' at which chaos propagates, to a general bound on the growth of chaos reminiscent of the ``Planckian'' bound on many-body transport.

Many open questions remain regarding the emergence of spacetime in the vicinity of black holes. A decade ago, serious obstacles were encountered in realizing a conventional description of the black hole interior in a microscopic theory compatible with the area law for black hole entropy. While some obstacles have been overcome, others persist, and it remains unclear to what extent the interior remains smooth for typical black holes. Relatedly, detailed understanding of the black hole singularity itself is still evolving. Current techniques for analyzing black holes on the gravitational side of the AdS/CFT correspondence remain approximate, leaving boundary observables that cannot be fully accounted for with existing bulk techniques. Addressing these questions requires a deeper formulation of quantum gravity than we currently possess, and comprises a clear goal for the coming decade.

\subsubsection{Gravity and the amplitudes program}

In recent years, it has become clear that deep issues in quantum gravity, including its surprising relation to gauge theories, can be understood through studies of scattering amplitudes. This is exemplified by the discovery and exploration of the double copy, which provides a means to calculate amplitudes in one theory using amplitudes from two technically simpler theories. Most prominently, this allows gravity scattering amplitudes to be constructed from pairs of gauge theory amplitudes. Such relations were originally discovered in string theory and subsequently greatly expanded by the discovery of a duality between color and kinematics, in which gravity amplitudes are generated by gauge theory amplitudes from the replacement of color factors with kinematic ones. This duality relates not only gravity and gauge theories, but also a wider variety of familiar field and string theories. Although first identified in perturbative scattering amplitudes, the double copy is being systematically extended to classical solutions as well. In the coming years, we can expect considerable development of the double copy in many directions.

Amplitudes methods have also contributed greatly to the understanding of the perturbative S-matrix in string theory. Recent developments have pushed string perturbation theory to three loops, leading to a number of non-trivial results. Future targets include circumventing technical obstacles to higher-loop string amplitudes and exploring the rich mathematical structure of Feynman integrals associated with K3 or Calabi-Yau geometries. Advances in string perturbation theory have enabled qualitatively new developments such as the Cachazo-He-Yuan (CHY) formalism and ambitwistor string models. The former expresses field theory amplitudes as integrals over the string worldsheet localized at the points which satisfy scattering equations; these integrals are built from simple building blocks and manifest connections between various field theories, including color-kinematics duality. The ambitwistor string expresses the same field theory amplitudes as the worldsheet correlators of ambitwistor strings, drawing fascinating new connections between string and field theory amplitudes. The existence of such novel descriptions of field theory scattering amplitudes in terms of string theory points to new underlying principles in quantum field theories.

\subsubsection{Particle physics and cosmology} 

In addition to deepening our understanding of quantum gravity itself, recent progress has also had far-reaching implications for particle physics and cosmology. Although the gravity is famously the one force not contained within the Standard Model, the Standard Model is ultimately embedded within a theory of quantum gravity. This motivates obtaining the Standard Model from a complete microscopic description such as string theory, in the process potentially addressing many of its outstanding mysteries. Progress has been most rapid in obtaining Standard Model-like theories with some degree of supersymmetry, inasmuch as this simplifies the construction of stable compactifications to four dimensions. In recent years, considerable progress has been made in the non-perturbative ``F-theory'' description of Type IIB string theory, in which supersymmetric Standard Model-like theories are obtained by engineering appropriate singularities in Calabi-Yau geometries. The classification of Standard Model-like theories translates into a concrete (albeit highly non-trivial) problem in algebraic geometry. Augmented by machine learning and other numerical developments, ongoing efforts are beginning to shed light on typical properties of these realistic theories, including the types of gauge groups, spectrum of charged states, and number of generations. A more ambitious goal is to obtain realistic models with supersymmetry broken at higher scales (such as the compactification scale or the string scale). This introduces new technical challenges, most notably the stability of scalar fields in the low-energy spectrum. Fruitful investigation is underway, and the construction of theories without low-energy supersymmetry is a compelling target for the coming decade.

Quantum gravity and cosmology are deeply intertwined. A fully quantum theory of gravity is necessary to understand the initial singularity of the universe, while the subsequent expansion of spacetime leaves imprints of short-distance physics visible on cosmological scales. Models of quantum gravity can inform both the early inflationary epoch and the current dark energy-dominated era. In string theory, the value of the cosmological constant depends on both the shape of the internal dimensions and the presence of additional extended objects such as branes and fluxes. Extensive progress has been made in constructing string configurations with positive values of the cosmological constant, fueling ongoing dialogue aimed at bringing these constructions under complete theoretical control.

A question of great experimental and theoretical interest is whether string theory or other frameworks for quantum gravity can produce inflationary models matching current and future observations. Of particular interest are predictions for the tensor-to-scalar ratio $r$, which is sensitive to the total displacement of the inflaton as it rolls down its potential and will be constrained at the level of $r \leq 10^{-3}$ by future experiments. Present constraints already exclude a range of effective field theories with large values of $r$, and it is a pressing question whether -- and where -- there is a theoretical upper bound on $r$ in complete models of quantum gravity. Microscopic models of inflation lead to a range of potentially observable features, including distinctive non-gaussian features in the cosmic microwave background or a reduced sound speed during inflation. A more complete theory of quantum cosmology is also essential to resolving the puzzles of ``eternal inflation'' that arise in many inflationary models. Eternal inflation in the semiclassical approximation gives rise to a vast, inhomogeneous universe where conventional notions of probability lead to bizarre paradoxes, the resolution of which is likely to be found in the theory of quantum gravity.

\subsubsection{Gravity and effective field theory}

The novel properties of a quantum theory of gravity provide new challenges and opportunities for effective field theory, both for gravitational EFTs and non-gravitational EFTs embedded in a theory of quantum gravity. 

Due to the presence of a massless graviton in the spectrum, gravitational EFTs have long resisted the development of positivity bounds akin to those arising in their non-gravitational relatives. However, rapid progress over the past few years has overcome many of the obstructions, leading to the first bounds on the operator coefficients of the leading corrections to Einstein gravity. In a complementary direction, CFT data has recently been used to constrain gravitational EFTs in AdS with holographic duals, including the establishment of rigorous bounds using the conformal bootstrap. 

In addition to the bounds on effective field theories arising from unitarity and causality, surprising constraints may also arise from consistent embedding in a theory of quantum gravity. These constraints are collected under the umbrella of what is now known as the ``Swampland program.'' Since a general definition of quantum gravity (and the universal properties of consistent theories coupled to it) are not rigorously established, these constraints are conjectural. However, in many cases they are supported by a diverse spectrum of evidence from perturbative string theory, semiclassical gravity, holography, and other sources. While some Swampland arguments can be related to positivity bounds of the types described above, in other cases they lead to entirely new constraints on EFTs. More broadly, phenomena such as black hole formation can lead to novel UV/IR infrared mixing phenomena that significantly generalize the standard notion of effective field theory and merit further study. Insofar as the Standard Model can be seen as an effective field theory coupled to gravity, these constraints and considerations provide a new avenue for addressing many of the Standard Model's long-standing puzzles.

\subsection{Broader implications for mathematics and physics} \label{sec:math}

In the course of answering questions in high-energy physics, fundamental theory uncovers both new mathematical structures and new approaches to problems in far-flung fields. 

From classical mechanics to general relativity, fundamental theories of physics have evolved hand-in-hand with mathematics. This is no less true for modern approaches to quantum field theory and quantum gravity. Recent implications for both geometric and algebraic aspects of mathematics have included direct connections between the classification of SCFTs and the theory of canonical singularities in algebraic geometry; wrapped brane states in string theory and the mathematics of enumerative invariants; and 2d rational CFTs and their associated vertex operator algebras. Connections between QFT, string theory, and number theory are exemplified by the ``moonshine program'' which has uncovered striking relations between modular forms and the representations of finite groups, disparate areas of mathematics that have been connected through the physics of 2d CFTs, vertex operator algebras, and associated constructions in string theory. Further relations between number theory, algebraic geometry, and quantum field theory are emerging from the geometric Langlands program. Recent work on perturbative scattering amplitudes also revealed deep connections to active areas of work in mathematics. This ranges from the theory of special functions, polylogarithms, elliptic functions, calculations of multi-dimensional integrals, and modern use of differential equations to intriguing links to combinatorics, algebraic geometry, Grassmannians and positive geometries. Exploring the field-theoretic implications of new mathematical structures, and vice versa, is bound to be fruitful for years to come.

Insofar as the AdS/CFT correspondence relates strongly coupled quantum systems to gravity, it can be used to understand a range of strongly-coupled interacting systems using straightforward computations on the gravity side. This has produced a number of insights in many-body physics including bounds on transport, bounds on chaos, and hydrodynamics with anomalies. In many cases, gravitational calculations inspired conjectures that were later proven in general, making the gravity side of the AdS/CFT correspondence a valuable laboratory for many-body physics. Strange metals (relevant for high-temperature superconductivity) have proven amenable to study using near-extremal black holes with a suitable horizon geometry, while models on the condensed matter side (such as the SYK model mentioned earlier) have improved our understanding of near-extremal black holes. Finally, qualitative similarities between the deconfined phase of QCD and maximally supersymmetric gauge theory at finite temperature have made the AdS/CFT correspondence relevant to the study of the quark-gluon plasma, leading to progress on real-time, out-of-equibrium problems including transport properties and jet quenching. 

Progress in quantum gravity has also driven progress in classical gravity. A good example is the recent use of quantum scattering amplitudes for pushing the state of the art in precision calculations of classical gravitational waves. Insights arising from the AdS/CFT correspondence include the discovery of the ``fluid-gravity correspondence'' connecting Einstein's equations to the Navier-Stokes equations of hydrodynamics, exotic new solutions to Einstein's equations, and astrophysical signatures of near-extremal Kerr black holes, as well as the derivation of the Penrose inequality (a proxy for cosmic censorship). Connections between soft theorems, the memory effect, and the Bondi, van der Burg, Metzner, and Sachs (BMS) group of asymptotic symmetries has been discovered, motivating new efforts to develop a theory of flat-space holography. Resonances between classical and quantum gravity are bound to grow in the coming years, with progress flowing in both directions.

\section{Phenomenology} \label{sec:pheno}

Particle phenomenology provides the connection between fundamental theory and the physical description of the real world, testable by experiments. Phenomenology has many crucial aspects essential for a successful experimental program in particle physics:
\vspace{-6pt}
\begin{itemize}\itemsep2pt\parskip2pt \parsep2pt
    \item  Phenomenology helps formulating the physics goals of particle experiments,  identifying promising new avenues for experiments, and also in interpreting their results, ideally by synthesizing them into a coherent model.
    \item Phenomenology provides the tools to perform the precision calculations necessary to compare the experimental results to theory predictions.
    \item Phenomenology helps providing many of the tools used for successfully analyzing and interpreting the results of the experiments. It often identifies signals requiring novel analysis techniques. 
\end{itemize}

Along these lines we distinguish the various aspects of phenomenology as model building, precision physics and collider phenomenology. We discuss cosmology and astrophysics as well as neutrino physics in separate subsections to highlight aspects of theory specific to these topics. 

Many aspects of particle phenomenology play a direct role in formulating the Projects of the next decade, and particle phenomenologists are actively contributing to or leading these studies. The role of the Theory Frontier is to highlight the important aspects of phenomenology that are providing more general tools or long term goals, which will be likely essential for the long-term success of the experimental program, but are not yet directly contributing to the Project-based frontiers. The direct contributions by particle phenomenologists will be highlighted in the Frontier Reports of the Energy, Cosmic, Rare and Precision, Computational and Neutrino Frontiers.  

\subsection{Model building} \label{sec:mod}

There are several clues indicating the need for physics beyond the Standard Model: the origin of neutrino masses, the quark and lepton flavor structure, the absence of CP violation in the strong sector, the coexistence of the weak and gravitational scales, the origin of dark matter  and dark energy responsible for the acceleration of the universe. Model building attempts to synthesize these clues into the next set of principles which determine the laws of physics at the shortest distance scales. While all BSM models approximately reproduce the SM at low energies, they introduce a plethora of testable ideas using a broad range of theoretical approaches and techniques. Many of these have led to new experimental search strategies at existing experiments, as well as to formations of wholly new experimental programs, often co-led by theorists.

\subsubsection{Naturalness} 

Naturalness has been one of the guiding principles for BSM model building over the past three decades. These naturalness puzzles arise in trying to understand why a dimensionless number is much smaller than ${\cal O}(1)$. These include the electroweak hierarchy problem ($m_h^2/M_{Planck}^2$), the cosmological constant problem ($\rho_\Lambda/M_{Planck}^4$), the strong CP problem ($\theta_{QCD}$) and the flavor problem ($y_f$).

The traditional approaches to the electroweak hierarchy problem are to extend the SM around the weak scale with new particles that have a built-in mechanism for eliminating further sensitivity to high mass scales. Generically these would involve new colored particles, in particular top partners. The canonical example is supersymmetry, which would predict scalar tops, or compositeness/warped extra dimensions which predicts fermionic top partners. These particles have been extensively searched for at the LHC without success, pushing the scale of such new particles beyond 1 TeV, leading to the reintroduction of some of the hierarchies (``little hierarchy") the models were intending to explain.  The status of weak scale supersymmetry is far from settled. In its original incarnation, in light of the LHC data, the MSSM now has to exist in moderately fine-tuned regions of the parameter space, but it is also possible that the superpartners most easily accessible are beyond the reach of the LHC such as in split SUSY, or that the structure of weak scale SUSY is qualitatively different from that of the MSSM. The search for warped extra dimensions/compositeness has been one of the driving forces of the experimental program over the past decade, leading to novel search strategies that were often applicable to other BSM scenarios as well. As for SUSY, the simplest models now have to live in moderately tuned regions, but again qualitatively different implementations are still possible and are continuously being explored and tested. 

One elegant way around the non-discovery of top-partners in BSM models explaining the electroweak hierarchy is Neutral Naturalness, where a new symmetry relates the SM quarks to colorless particles.  The most well-known of these is the Twin Higgs model, and a generic feature is the emergence of multiple sectors related by some discrete symmetry. These models typically require UV completions with colored states at the multi-TeV scale, motivating future high-energy colliders. The Higgs often acquires new or exotic decay modes, proportional to the SM Higgs to the twin Higgs VEV $v/f$, which is also directly proportional to the tuning in the model. The exact implementation of neutral naturalness will significantly influence the resulting phenomenology. Particularly interesting is the case where the twin confining sector has no light quarks, leading to a ``quirky phenomenology", where heavy quarks behave as if connected by a string with constant tension, giving rise to a shower of glueballs, some of which have displaced decays to SM states. Neutral naturalness is expected to be one of the dominant paradigms driving experiment and theory over the next decade. 

An important new direction that emerged over the past decade is the cosmological selection of the electroweak vacuum, which was inspired by similar attempts at selecting the cosmological constant or $\theta_{QCD}$. The best known example is the relaxion, where an axion-like field (the relaxion) traverses large values in field space, and provides a rolling Higgs mass in addition to the usual axion-like coupling to $G\tilde{G}$ of QCD. Once the Higgs develops a VEV wiggles will develop for the relaxion due to the usual QCD-like contribution to the axion potential, which will stop the rolling of the relaxion. Building a realistic model is challenging, in particular it is very hard to protect the shift symmetry of the axion over trans-Planckian field excursions. Another popular model of cosmological selection is called N-Naturalness, containing a very large number of copies of the SM, but only those with a small Higgs mass will acquire significant energy densities after reheating. Cosmic selection models are expected to be among the dominant topics in the near future, stimulating new experimental searches and predicting novel signals. 

The most famous approach to the strong CP problem is the QCD axion, which dynamically adjusts its VEV to cancel the strong CP phase. Search for the axion has motivated a thriving experimental program with lots of synergy between theory and experiment, which in turn has rejuvenated axion model building, where the properties of the QCD axion are varied away from the naive expectations. One challenge axions are facing is the quality problem: to ensure there be no sources of explicit breaking of the shift symmetry other than QCD itself, which has inspired another branch of axion model building. Axions can also play a role in large-field inflationary models. Axion physics will likely keep attracting enormous attention and lead to new experimental setups for axion searches.

\subsubsection{Dark matter} 

The existence of dark matter (DM) is one of the most compelling pieces of evidence for physics beyond the Standard Model. Dark matter has driven many of the developments over the past decade in particle theory, particle astrophysics and cosmology. It has also been an area of major focus for model building, with null results at large direct detection experiments driving increasing efforts to explore the vast theory space for dark matter. Insofar as dark matter model-building is inextricably linked with theory-led proposals for dark matter detection, the main discussion of the topic is reserved for Sec.~\ref{sec:cos}.

\subsubsection{Baryogenesis} 

The observed baryon-antibaryon asymmetry in our Universe calls for a mechanism of baryogenesis after the reheating post-inflation. Any such mechanism has to satisfy the three Sakharov conditions of baryon number violation, C and CP violation, and departure from equilibrium. While the SM nominally satisfies all three, the CP violation due to the phase in the quark mixing matrix (CKM) appears to be too weak, and no processes in the SM go out of equilibrium in the early Universe, necessitating BSM physics for baryogenesis.  The best-known model is leptogenesis, where a leptonic asymmetry generated by right handed neutrino decays is transferred to a baryon asymmetry via electroweak sphalerons. Another option is electroweak baryogenesis, happening during the electroweak phase transition. There have been numerous novel models introduced over the past decade, resulting in many experimentally testable models, since they involve low scales. This topic is expected to keep attracting ever growing attention. 

\subsubsection{Flavor and neutrino models} 

Flavor violating processes, in particular those involving flavor-changing neutral currents (FCNCs), have exquisite sensitivity to new sources of flavor and CP violation, due to the special flavor structure of the SM. The origin of this flavor structure including the hierarchical quark and lepton masses, the hierarchical CKM matrix, and the non-hierarchical neutrino mixing matrix (PMNS) is referred to as the SM flavor puzzle. Various classes of models exist to solve it, including horizontal flavor symmetries, warped extra dimension/partial compositeness, and radiative fermion masses. Quark FCNCs can often indirectly explore mass scales well beyond the reach of current colliders, while charged lepton FCNCs and electric dipole moments are powerful null tests of the SM. Over the last several years a number of flavor anomalies have emerged creating considerable excitement in the community. These include some decay modes of the $B$ meson as well as the anomalous magnetic moment of the muon. If these anomalies survive they would have transformative impact on particle physics, establishing a new mass scale, providing a new target for direct exploration at future colliders.

\subsection{Precision physics\label{Sec:Precision}} 

Theoretical techniques for precision physics are the backbone of a successful program in particle physics.   They are necessary for determinations of fundamental parameters of the Standard Model (SM) to unprecedented precision and probe beyond SM (BSM) physics to very high scales. Potential dark matter candidates, explanations of the hierarchy puzzle, and many other discoveries could be made, even if the mechanisms underlying them are at the tens of TeV scale or even beyond. Here we review some of the recent advances and expected future prospects for precision physics appearing both for collider phenomenology as well as for flavor physics. 

\subsubsection{Precision collider phenomenology}

The first step is the calculation of the perturbative hard scattering cross sections to the requisite order in perturbation theory. There has been tremendous recent progress on this front,  mostly due to a cross-fertilization between
formal/mathematical developments in understanding the properties of multi-loop amplitudes and to a clever
and efficient reorganization of different contributions into infrared (IR) safe observables up to next-to-next-
to-leading order (NNLO). As a result, the current standard for cross section predictions at the LHC is next-to-next-to-leading order (NNLO) in perturbative QCD for $2\to 2$ scattering processes without internal mass scales
in the contributing loops. For a host of processes the N$^3$LO corrections in perturbative QCD are
also available. The challenge facing the theory community is to extend these computations of
hard scattering cross sections at N$^3$LO to include jet processes and more differential observables. Numerous issues must be addressed to achieve this challenge: the computation of the relevant
three-loop integrals, understanding the basis of functions needed to describe these corrections, and
the extension of infrared subtraction schemes to handle triply-unresolved limits. Moreover, higher-order electroweak (EW) corrections become increasingly important with improved precision of LHC analyses. NLO EW corrections for $2 \to n$ processes can now be carried out in a mostly automated fashion (for $n$ up to 8). A few results are already available for NNLO mixed EW-QCD corrections, while only one  NNLO EW result (for the leading $Z$-resonance contribution to Drell-Yan production) has been obtained so far. The computation of higher-order EW corrections involves unique challenges due to the appearance of many masses in the loop integrals, which impedes analytical solutions to the latter.

While computations of the hard scattering cross section at fixed orders in perturbation theory are
sufficient for many applications, this is not the case for all situations. In multi-scale problems, when one mass scale is very different than the others, large logarithms of the scale ratios can appear,
and fixed-order perturbation theory doesn’t converge. To get a reliable prediction one must resum
these large logarithms to all orders. There have been many important new devlopments on this front, including resummation beyond leading power, the joint resummation of
different classes of logarithms relevant for jets and their substructure, small-x resummation in the high-energy
regime, and the QCD fragmentation processes.

All hadron collider predictions require understanding of the parton distribution functions (PDFs) that
describe how to take a parton of a given momentum fraction from a hadron.  
Although there has been significant recent activity in attempting to calculate PDFs using lattice
techniques, for the practical purpose of predicting collider physics cross sections they are
extracted from experimental data. This leads to the following three issues that must be addressed
in order to have PDFs computed to the needed level for LHC and future collider predictions:
the perturbative DGLAP (Dokshitzer–Gribov–Lipatov–Altarelli–Parisi) evolution of the PDFs must be calculated to match the precision of
the hard scattering cross section; 
the hard scattering cross sections from which the PDFs are
extracted from data must be known to NNLO or N$^3$LO depending on the desired accuracy; the
experimental data used in the extraction must have small enough uncertainties to match the
above theoretical uncertainties, and must also be sufficiently broad enough to fix the functional
dependencies of all PDFs on $x$.

Both fixed-order and resummed predictions describe collider observables that are inclusive, or
differential in a few variables. They are formulated in terms of partonic degrees of freedom.
Parton-shower event generators are needed to provide a closer realization of the actual events in
terms of hadronic degrees of freedom measured in experiment.  Parton-shower event generators such as HERWIG, PYTHIA and SHERPA
are heavily used by experimental collaborations in their analyses and form an indispensable tool
for understanding events at high-energy colliders. There have been numerous theoretical improvements in parton showers over the past years, resulting in programs more faithful to the underlying QCD theory. Future precision measurements of the Higgs and electroweak sector at planned Higgs factories will require N$^3$LO and possibly even N$^4$LO EW and mixed EW-QCD corrections calling for new innovative techniques that combine analytical and numerical methods, while planned multi-TeV proton-proton and muon colliders will require novel tools for the resummation of multiple emission of massive gauge and Higgs bosons (EW parton distribution functions, EW parton showers, etc.) and energy-enhanced radiative corrections.

The SM effective field theory framework will inevitably be the framework in which future indirect searches for physics beyond the SM are interpreted. 
The SMEFT Lagrangian is constructed containing only the SM degrees of freedom, and assuming that all operators satisfy the SM gauge symmetries. This leads to a result that differs from the SM Lagrangian by a series of higher-dimensional operators suppressed by a high energy scale $\Lambda$ at which the EFT description breaks down. Predictions in this EFT require not only expansions in the usual SM couplings, but also in the ratios $v/\Lambda$ and $E/\Lambda$, where $v$ denotes the Higgs vev and $E$ the characteristic energy scale of the experimental processes under consideration. As experimental precision increases, and as $E$ increases, higher operator powers must be considered for reliable predictions. An essential step for interpretation of future high-precision HL-LHC data within SMEFT is the extension of predictions beyond the leading dimension-6 order. It has been demonstrated that constraints derived assuming only dimension-6 operators can be unstable against the inclusion of dimension-8 effects. It has also been shown that dimension-8 terms are needed to faithfully reproduce the underlying UV theory in some cases. Significant effort must be devoted to understanding the impact of dimension-8 and higher-order effects.

\subsubsection{Precision Flavor Physics} 

A key feature of flavor physics — the study of interactions that distinguish between the three generations — is the plethora of observables that probe very high mass scales, well beyond the center-of-mass energy of the LHC or any planned future colliders.  This high mass-scale sensitivity arises because the SM flavor structure implies strong suppressions of flavor-changing
neutral-current processes (by the GIM mechanism, loop factors, and CKM elements). The ongoing and planned flavor experiments provide essential constraints and complementary information on BSM models. The measurements of dozens of CP-violating and FCNC processes at $e^+e^-$ colliders and at the LHC are consistent with the SM predictions, with ever-increasing precision (with a few exceptions; see the discussion of flavor anomalies below), leading to the “new physics flavor problem”, which is the tension between the hierarchy puzzle motivating BSM physics near the electroweak scale, and the high scale that is seemingly required to suppress BSM contributions to flavor-changing processes. 
The bound on the scale of flavor violating dimension-6 operators reaches above $10^5$ TeV for the most sensitive observables like $\epsilon_K$ and the electron EDM, with the greatest improvements
over the next decades expected in $\mu \to e$ conversion ($\mu N \to e N$) and EDM experiments. These bounds also have important impact on model building: for SUSY one needs to augment the model with some mechanism to  suppress the SUSY flavor violating contributions, 
for example via degeneracy, quark-squark alignment, heavy (3rd generation) squarks, Dirac gauginos with an extra R symmetry, or split SUSY. Warped extra dimensions have a built-in RS-GIM mechanism that naturally reduces FCNC close to the experimental bounds. A widely applied concept in BSM model building is minimal flavor violation (MFV), which posits that BSM sources of the breaking of the global $[U(3)]^5$ symmetry of the SM are proportional to the same Yukawa couplings, which often reduces FCNCs to lie within the bounds. 

The richness of $B$ physics and the large $b$-quark mass enable many complementary tests of the SM, and have been a driving force to develop new perturbative multi-loop and nonperturbative effective field theory techniques since the 1980s. For example, a precise determination of $B \to X_s\gamma$
makes it essential to consistently combine fixed-order perturbative calculations, resummations
of large logs (of various kinds) in endpoint regions, and nonperturbative ingredients, each of which has grown into significant areas of research. The desire for (hadronic) model-independent understanding of semileptonic $B$-meson decays led to the development of heavy quark effective theory (HQET), which was instrumental in the development of many other effective field theories including nonrelativistic general relativity (NRGR). Studying inclusive $B\to Xl\nu$  decays led to the development of the heavy quark expansion, which has served as a model for other operator product expansions. The soft-collinear effective theory was developed initially motivated by summing Sudakov logs in $B \to X_s\gamma$ in an effective theory framework. SCET has also become part of the standard theory tool-kit for higher order collider physics calculations. In the area of multi-loop calculations, flavor physics also instigated developments of numerous technical aspects. This decade Belle-II is expected to produce new data for inclusive $B$-meson decays such as $B\to X_s\gamma$ and $B\to X\ell \nu$. To match this experimental development, new theoretical work that utilizes HQET and SCET is needed. It will combine existing and new perturbative calculations as well as better control of non-perturbative inputs.

Over the past five years several anomalies have also emerged from the flavor experiments, that could potentially point to new physics BSM if these anomalies persist after further experiments and more precise theoretical determination. Most prominent among the anomalies are the hints for lepton flavor universality violation (LFUV) in two set of processes. One is in neutral current $b\to sl^+l^-$ transitions, $R_{K^{(*)}} = {\cal B}(B \to K^{(*)}\mu^+\mu^-) /{\cal B}(B \to K^{(*)}e^+e^-)$, where the most recent LHCb measurement shows a 3.1$\sigma$ deviation from the SM,  while the other is in the charged current $b\to cl\nu$ transition, $R(D^{(*)}) = {\cal B}(B\to D^{(*)}\tau\nu)/{\cal B}(B \to D^{(*)}l\nu )$, where $l = e,\mu$.   In this case the significance for the deviation from the SM is quoted as 3.1$\sigma$  to 3.6$\sigma$, depending on the treatment of correlations. In both of these cases, the present data hint at about $15-20$\% corrections to the SM predictions. 
In the SM predictions of the $R_{K^{(*)}}$ ratios, contributions from form factors and non-factorizable effects largely cancel, yielding theoretical predictions with negligibly small  uncertainties. Hence more precise measurements of $R_{K^{(*)}}$ could break the SM (if the central values stay unchanged). SM predictions of the $R(D^{(*)})$ ratios, on the other hand, depend on the underlying form factors, which must be known with sufficient precision in order to assess the tensions between the SM and experiment. Lattice-QCD results for semileptonic $B$-meson decay form factors are currently available with complete error budgets at a commensurate precision level to experiment, also yielding reliable determinations of CKM elements from experimental measurements of differential decay rates. To take advantage of more precise experimental measurements, there are ongoing efforts to reduce the uncertainties in lattice-QCD form factor calculations in tandem (see Sec.~\ref{sec:lft}).    
With the expected improvements in experimental precision, a common challenge in both $b \to ql\nu$  and $b \to qll$ mediated decays is that going forward, the role of electromagnetic corrections needs to be better understood, and the related uncertainties assessed.  

To fully utilize the next generation of measurements, a better theoretical understanding of nonleptonic decays is much desired. To date, most CP violation measurements have been performed for such decays, as they not only allow measurements of the CKM unitarity triangle angles, but also provide numerous probes of CP-violating BSM interactions. A promising recent theoretical development extends factorization of nonleptonic decays to order $\Lambda_{QCD}/m_b$,  with regularized endpoint divergences. This can lead to a better theoretical understanding and control of nonleptonic decays. There is also active ongoing research on studying CP violation in charm and kaon decays as probes of the SM and new physics.  

\subsection{Collider phenomenology} \label{sec:coll}

Collider phenomenology is a highly active theoretical area that connects in crucial ways to
the experimental energy and precision frontier programs. It is an essential interface between the theoretical and experimental high-energy physics communities, serving various roles: from connecting formal investigations to experiments, such as providing guidance on the exploration of the physics possibilities, to supporting the experimental community with the essential tools to simulate and interpret data, bringing information back to the theory community. Below we review some of the most exciting recent developments  as well as the directions with most promise for the coming decade. 

\subsubsection{Collider observables}

At the heart of collider phenomenology is choosing and defining observables. While cross section is the most basic observable, it doesn't contain much information about the microscopic physics. By making more differential measurements, one gains access to different regions of phase space
and rarer processes, and therefore increases the information. The design of observables also plays a key role
in separating out known background processes from the signals of interest. Often it is important to develop observables that expose kinematic features, leading to the development of specialized observables  which are expected to play an important role in studying new physics scenarios with multi-body
final states. An area of rapid development over the past decade has been the field of jet substructure: variables that are designed for tagging jets either as arising from SM partons, or exotic jets arising from new physics scenarios. A newly emerging and very promising field is the study of multi-point correlators, where each event contributes to several entries of the histogram, weighted by the product of particle energies. Because of this weighting, energy correlators are less sensitive to soft physics and can be calculated accurately. In addition the scaling behavior of these correlators can often be deduced from the conformal limit. We expect these novel theory driven observables to play a major role in the collider physics of the next decade. Another new method revolutionizing the subject of collider observables is using machine-learning (ML) based observables, where data itself is used to define the observables. One important new insight is that ML can incorporate symmetry
group equivariance, such that these methods respect the know theoretical structures of collider physics,
such as permutation and Lorentz symmetries. An important example of unsupervised ML is optimal transport, which is based on the relationship between pairs of collider events, rather than treating them independently. This method led to strategies for quantifying the dimension of collider events, and to new observables like event isotropy. Another emerging direction is the use of quantum algorithms for collider physics. It has been successfully used to compute collider observables; whether they will eventually lead to a dramatic computational gains depends on whether efficient ways for encoding classical data in quantum form can be found.

\subsubsection{Precision calculations}

Once an observable has been defined, one needs to calculate the distribution accurately enough
in the Standard Model to match current and expected precision from experimental measurements. In
addition, predictions in specific BSM scenarios or in effective field theories are
needed to establish the sensitivity to new physics as well as to devise optimal search strategies. We are reviewing the recent progress and future prospects of the precision collider physics calculations in Sec.~\ref{Sec:Precision}.

\subsubsection{Event generators}

Very few theoretical calculations can be performed analytically, so Monte Carlo event generators are the workhorse strategy to make theoretical predictions which can most directly be compared to experimental data.
As experimental methods become more sophisticated, theoretical innovation in multi-purpose event
generators becomes increasingly important. In precision collider measurements, a significant (or even dominating) source of uncertainties of experimental analyses is often associated with event generators. A large part of recent development activities in Monte Carlo event generators focuses on extending
their applicability and reducing their uncertainties. Event generation at present and future colliders feature many common ingredients: higher-order QCD and EW corrections, factorization theorems and parton evolution, resummation of QCD and QED effects, hadronization, and initial and final-state modeling. In addition to the
physics components, there are also computing elements, such as interfaces to external tools for analysis,
handling of tuning and systematics, and the need for improved computational efficiency. In addition to being a theoretical challenge, the increase in precision is becoming a computational challenge, partly because the complexity of higher-order predictions for multi-particle final states increases exponentially with the order and number of final states, and partly because the expected increase in collected data calls for the production of very large sets of fully simulated events. One way of address the ballooning computational cost of event generators is through machine-learning-based generators.  Modern machine learning is driving recent progress in event generation, simulation, and inference for high-energy colliders.

\subsubsection{Interpretation tools}

Interpretation lies at the intersection of theoretical and experimental collider physics, which requires dialogue between communities, and collider phenomenology is a language to facilitate that dialogue. Model-specific searches for new physics are a well-understood strategy for data interpretation, but there is increasing interest in anomaly detection for collider physics. A quasi-model-independent approach to interpretation is the use of effective field theories. If the impact of heavy new physics states can be captured by contact interactions involving Standard Model fields, then one can do a systematic expansion order by order in power counting, leading to the Standard Model effective field theory. A key challenge facing collider physics is data and analysis preservation. To maximize the scientific potential of archival collider data sets, theoretical physicists should actively participate in proposing and stress-testing archival data strategies.

\subsubsection{Search strategies} 

There is a huge range of possible scenarios beyond the Standard Model, and each scenario requires different observables to maximize signal acceptance and minimize background contamination. Model-specific searches remain the gold standard for the field, since they provide a well-defined statistical framework for setting limits (or announcing a discovery). 
Theorists have made significant contributions to proposals of novel measurements of the newly discovered 125 GeV Higgs boson, including for example exotic Higgs decays, measurement of the Higgs width and the Higgs coupling to charm. 
Model agnostic searches, though, are gaining traction, as it becomes increasingly possible to automate certain aspects of the search process. The most heavily studied examples include the cascade decay signature appearing in models with an extended Higgs sector, composite Higgs models with vector-like fermions or warped extra dimension models with Kaluza-Klein (KK) modes. Heavy particles produced at high-energy colliders could lead to boosted objects with interesting substructure. 
The study of dark sectors has been a subject of interest in recent years, including dark showers, portal matter, and multi-field scenarios. The field of detecting long lived particles (LLPs) has also undergone a major revolution: besides new analysis techniques at existing experiments several auxiliary detectors like FASER, Mathusla or CODEX-b have been proposed and co-led by theorists, some of which are already under construction.  Low-mass scalars are a key target for current and future colliders, including axion-like particles and light scalars from extended Higgs sectors. 

\subsection{Cosmology and astrophysics} \label{sec:cos}

Cosmology and astrophysics provide a wide range of opportunities to expand our knowledge of the fundamental laws of nature, both through direct searches for BSM physics and through tests of the SM in extreme conditions that are impossible to recreate in the laboratory. The impact of theoretical effort in cosmology and astrophysics over the past decade can be viewed through the lenses of (i) advancing our understanding of fundamental physics by forcing us to ponder extreme scenarios where, e.g., quantum effects and gravity must be considered simultaneously, (ii) developing new microscopic models that can potentially explain the outstanding problems facing our understanding of nature, and (iii) inventing new approaches to test our best-motivated theories, in addition to developing the theoretical tools needed to properly interpret the resulting data.

\subsubsection{Dark matter}

We have overwhelming evidence for the existence of cold and at most weakly interacting DM across a
variety of astrophysical scales comprising $\sim 27\%$ of the energy density of the universe. The DM mass is constrained on the lower end to be larger than $\sim 10^{-20}$ eV and on the upper end to be smaller than $\sim 10^4 M_{\odot}$. We find ourselves at a unique point in history, where the existence of DM on astrophysical
and cosmological scales is known and well characterized, but the microscopic nature of the DM
is largely unconstrained, up to rough constraints on the DM mass and interaction
strengths with itself and with ordinary matter. Over the past decades theorists have developed particle-physics-based models to explain DM that typically lead to faint but observable signatures. The types of signatures vary drastically depending on the DM model
at hand, leading to a world-wide scientific program searching for evidence of particle DM
across laboratory experiments ranging from the LHC, to underground direct detection experiments, to precision laboratory experiments, to intensity frontier experiments. In many DM models there are direct or indirect astrophysical signatures, such as modifications to DM halo structure itself; electromagnetic, cosmic ray, and neutrino signatures in the case of annihilating, decaying, and converting DM; and even gravitational wave signatures associated with phase transitions in the early universe. DM may also be associated with new light physics, such as light mediators or ultralight
particles, that may have effects ranging from modifying stellar evolution, to producing signatures in dedicated fixed target experiments, to adding extra radiation measured in the cosmic microwave background (CMB). Theoretical efforts to produce viable DM candidates have been the driving force behind numerous
experimental and observational programs. Beyond
the construction of specific DM models, theorists have played an especially important role in the
past decade in proposing and helping implement specific experiments and astrophysical search
strategies for covering some of the best motivated DM parameter space.

Weakly interacting massive particles (WIMPs) with electroweak scale masses
and couplings that acquire their relic abundance through thermal freeze-out in the early universe have been considered the canonical DM candidates for decades, partly because  solutions to the hierarchy problem  (such as low-scale SUSY) naturally produce WIMP-like DM candidates. However, at present the search for WIMP DM is at a turning point, since both direct and indirect searches for SUSY and other TeV-scale models of naturalness have so far yielded no direct evidence, tightly constraining large areas of the parameter space of these models as well as  the properties of a putative WIMP DM candidate.  On the other hand, WIMP DM is very much not dead. Nearly-pure Higgsino DM serves as an illustrative example of a surviving WIMP DM candidate that serves as an exceptionally well-motivated target for upcoming experiments. Composite Higgs and Twin Higgs models also contain a variety of well-motivated DM candidates.  One area with substantial theoretical effort and progress over the past decade has been the study of indirect detection of WIMP-like DM via annihilation or decay into high energy photons in the X-ray or $\gamma$-ray band. Theorists have been key parts of this effort, and have developed new analytical strategies and analysis tools for these searches. Progress in experimental tests of these two landmark signatures has been accompanied by a substantial broadening of astrophysical searches for the imprint of DM in cosmic rays.

The past decade has also seen enormous progress, led by theoretical efforts, in constructing models of particle dark matter that go beyond the WIMP paradigm. A notable variation is where the DM particle itself does not interact significantly with the SM, but instead it interacts with a secondary state, which itself has a standard freeze-out mechanism. If the mass of the secondary state is much heavier we get a secluded model with strongly suppressed direct detection rates.  If instead the secondary state's mass is close to the DM mass we obtain a co-annihilation model.   One popular class of models is based on the realization that $3\to 2$ or $4\to 2$ interactions (a phase of ``cannibalism") can substantially influence the resulting DM relic density. Resulting models include SIMPs, co-SIMPs and ELDER. Another mechanism for generating DM is freeze-in. Here the dark sector is populated by the leakage of energy from the visible sector through sub-Hubble annihilations or decays of SM particles over time. These models (often also called FIMPs) include sterile neutrinos, singlet scalars, and various superpartners. Alternately, the observed dark matter relic abundance may be generated from a residual asymmetry. For example, the concept of asymmetric dark matter posits a common origin for the baryon content of the Universe and DM. An interesting mechanism for enhancing the late-time annihilation signal is the Sommerfeld enhancement, which happens when long range forces arise in the non-relativistic limit. Another important dynamical concept is Inelastic DM, when the scattering of DM on nuclei can only happen via an inelastic channel requiring a minimal threshold energy for the incoming DM particles, potentially suppressing certain direct detection cross sections. It is also possible that the DM is a stable bound state of a confining sector, where some symmetry guarantees the stability of this particle. They could either be dark pions, or more naturally dark baryons, or even dark glueballs. Another interesting possibility is atomic DM, formed by atom-like bound states of some hidden/mirror sector, most commonly two fermions oppositely charged under a $U(1)_D$ dark gauge symmetry.

Perhaps the most striking progress in dark matter theory over the past decade has come from renewed attention to axionic, light, and ultralight DM. The QCD axion was originally introduced to address the strong-CP problem, but axion-like particles which do not couple to QCD are also motivated for purely theoretical reasons because they arise generically in the context of string theory constructions.  They can also be very successful DM candidates, with their dynamics depending on the relation between the Peccei-Quinn (PQ) symmetry breaking scale and the scale of inflation. For a high PQ breaking scale one needs dedicated lattice-QCD calculations to determine the axion DM abundance, while for a low PQ breaking scale one needs dedicated simulations of the production of axion strings and domain walls. Such simulations have seen a revolution in their complexity and accuracy in the last decade, aided in large part by advances in high-performance computing, and these computations will improve further in the near future by leveraging computational and technological advances. Axion-like particles as well as a broader class of scalar and vector particles  may also make up a sizeable fraction of the DM. 
Over the past decade ``fuzzy DM" --- ultralight dark matter with wavelength on astrophysically relevant scales--- has received a wave of interest in part to explain a number of apparent failings of the standard cosmological model on small astrophysical scales. On the other hand, fuzzy DM has become increasingly constrained in the latter half of the past decade, thanks to theoretical efforts in understanding the astrophysical implications of fuzzy DM, cosmological and galactic-scale simulations incorporating fuzzy DM, and new data. 
It has long been understood that axions could leave detectable astrophysical signatures by modifying stellar cooling. The recent refinement of stellar and compact object cooling probes and the development of a number of novel astrophysical tests have combined to provide some of the strongest constraints on the QCD axion and ultralight scalars, vectors, and axion-like particles, apart from a narrow mass range probed by the ADMX experiment.

Traditionally, direct searches for axions fell into one of the following three classes:  (i) light shining through wall experiments, (ii) axion helioscopes, and (iii) axion resonant microwave cavity haloscopes. Thanks to pioneering theoretical effort, followed by bold and innovative small-scale experimental programs, there now exists proposals for probing nearly the entire currently-allowable QCD axion mass range. An important theoretical observation at the beginning of the last decade was that the axion, while it solves the strong-CP problem by removing the time-averaged neutron EDM, leaves a time-varying, residual EDM when the axion is DM. This led to the proposal for the CASPEr experiment, which aims to detect the oscillating axion-induced EDM using a nuclear magnetic resonance based experiment.  A similarly ground-breaking discovery came from the observation that by thinking more broadly about the modifications to Maxwell’s equations in the presence of axion DM, experimental setups could in principle be constructed that would be sensitive to GUT scale axion DM using the axion-photon coupling. This became the basis of the ABRACADABRA 10-cm collaboration, whose initial demonstrator has already set world-leading limits on the axion-photon coupling.  Other novel proposals in ultralight dark matter surround the idea that scalar dark matter coupled to the Standard Model can result in small periodic oscillations of fundamental constants such as the electron mass and the fine structure constant.  Searches performed to date have taken advantage of precision measurement techniques such as atomic clocks to search  new parameter space.

The indirect detection of DM annihilation or decay in the local universe through its imprint on cosmic rays is a key element of the program to detect and identify DM. In recent years, indirect detection analyses have made major progress in exploring the parameter space where signals are predicted in the most straightforward models, such as WIMP DM and sterile neutrino DM. Theorists have been key parts of this effort, and have developed new analytical strategies and analysis tools for these searches. Progress in experimental tests of these two landmark signatures has been accompanied by a substantial broadening of astrophysical searches for the imprint of DM in cosmic rays. Recent developments in DM model-building have expanded the range of DM annihilation and decay signatures that are of interest to indirect detection, providing models with novel spectral signatures, signatures that populate new kinematic areas, signatures in novel final states, and signatures with unusual spatial distributions. It has long been understood that axions could leave detectable astrophysical signatures by modifying stellar cooling. However, in the past decade a number of novel astrophysical probes of
axions and axion DM have been developed, in part because of input from string theory motivating axion-like particles and broader parameter space for the QCD axion. At the same time, the stellar cooling probes have been refined, such that at-present the strongest constraints on the QCD axion
and axion-like particles, apart from a narrow mass range probed by the ADMX experiment, arise from astrophysical probes.
Theorists have been crucial in developing the ideas behind these searches and implementing them with astrophysical data.

Another key area where a new approach to DM model building and detection has seen rapid progress is that of sub-GeV dark sectors, which contain new singlets under the SM gauge symmetry that may or may not have self interactions. If DM resides in the dark sector it can talk to the SM through a number of weakly-coupled portals. Dark sector constructions often
contain light mediators, such as dark photons, that can themselves be directly probed by experiments. Theorists have not only developed models of sub-GeV dark sectors, but also played important roles in devising new approaches to search for them, from new types of direct detection to accelerator production of dark matter and accelerator searches for the
mediators themselves. 

Prior to the early 2010’s, the experimental effort for seeking the direct detection of DM was singularly focused on elastic or inelastic scattering off of nuclei. However, at the beginning of the past decade, motivated by sub-GeV dark sector modeling efforts, a range of theoretical proposals was put forth for detecting novel inelastic DM interactions with electrons in materials. The advantage of scattering off of electrons for light DM is clear: since the electron has less mass, it acquires a larger recoil energy relative to nuclei during scattering processes. Since electrons tend to be bound, DM-electron processes tend to be inelastic.  It was also proposed that semi-conductor targets might be even better suited for light-DM searches, since the DM would in principle just need enough energy to push an electron over the band-gap in order to produce a detectable signature, though calculating the scattering rates is non-trivial and connects with cutting-edge topics in condensed matter physics. The theoretical work showing the promise of semiconductor targets for light DM motivated an experimental research program to try to detect one or few electron events. The detection of single electron scattering events is now possible, as demonstrated by the SuperCDMS and SENSEI collaborations.

A second path to detecting  sub-GeV dark matter is through its production at accelerators. Unlike direct detection, where light-DM signals are kinematically challenging to detect, the energy required to produce light DM at accelerators is achievable at many facilities, including both flavor factories and fixed target experiments. Instead, the challenges of
accelerator-based detection stem from the very weak couplings typically expected for light DM. Theoretical studies were the first to note that, despite these small couplings, accelerator-based neutrino experiments had sufficiently high luminosity to produce a secondary beam of
DM particles, leading to a potentially observable scattering rate in downstream detectors. This realization motivated the MiniBooNE-DM experiment, as well as searches for DM at the COHERENT and CCM detectors.  A second approach to searching for light DM at accelerators, the missing energy/momentum strategies, leverages the distinctive kinematics
of DM production reactions to identify them with high efficiency in a lower luminosity lepton beam.  Theorists again played important roles in developing these strategies, exemplified by NA64 and LDMX, and understanding their ultimate capabilities.

The realization that DM may reside in a dark sector weakly coupled to the SM has also led to new pathways to discovery for the mediator itself at accelerator-based experiments. The idea of looking for weakly-coupled mediators is distinct from the traditional approach to new-physics searches at colliders, which tend to focus on the energy frontier and probing new, relatively
strongly coupled, but massive states. This led to a
number of theoretical proposals for dedicated or parasitic dark-sector collider-based searches
over the past decade, many of which have already turned into actual experiments or are funded
and in construction phases, for example HPS, FASER, APEX or DarkQuest.  
 
\subsubsection{Quantum sensing}

As discussed in the preceding subsection, there is a plethora of examples where theorists have instigated new DM experiments. The development of quantum sensing technologies, in particular, has opened the door to new opportunities to search for new particles or interactions that arise in well-motivated BSM theories, enabling novel experiments, detectors, and measurements. Here high energy theorists are playing a central role in proposing and guiding such experiments in the pursuit of searches for new light matter particles (axions, axion-like particles, dark photons, milli-charged particles), dark matter, gravitational waves over a large range of frequencies, and tests of quantum mechanics and of gravity -- the list is long. The envisioned experiments employ diverse quantum sensing platforms and technologies, including atom interferometers, atomic clocks, novel condensed matter systems, SRF cavities, dielectric stacks, dish antennas, quantum optics, quantum spin gyroscopes, and single particle traps among others. They typically include interdisciplinary efforts connecting theorists and experimentalists from HEP, AMO, and condensed matter. As proposals are being turned into experiments, theorists continue to provide guidance and interpretation, strengthening the connections between theory and experiment to the benefit of both.

\subsubsection{Cosmology} 

Cosmological observations offer the unique opportunity to reconstruct the history of the universe
and the laws that shaped it.  It is through
our theoretical understanding of the forces that shaped the cosmos that we can then reconstruct
the expansion history, search for new particles and forces, identify new objects, and more.

The cosmic microwave background has been the main driving force in cosmology to date. The impact of new CMB data will depend increasingly on theoretical techniques to isolate different physical effects that alter the observed high-precision CMB maps. CMB photons are both gravitationally lensed and scattered by the matter
between us and the surface of last scattering. Ongoing theoretical work has shown how these
effects can be removed from the CMB maps using their statistical properties and frequency
dependence, in the process creating new maps for the distribution of matter in the universe. Galactic foregrounds, particularly from dust, present an additional challenge to CMB measurements. Our understanding of these dust signals from first principles is limited but has been bolstered by simulations and data-driven techniques. 

Gravitational waves produced after inflation are an increasingly compelling window into the
history of the universe. This includes both gravitational waves produced during reheating and/or
phase transitions in the early universe and from mergers of black holes, neutron stars, and/or
more exotic objects. Given that the universe is transparent to gravitational waves, this
presents a unique opportunity to probe important events throughout cosmic history and not just after recombination. Generating templates at the accuracy needed for future observatories will require a
variety of theoretical tools that connect observational needs to the most fundamental questions
about the nature of gravity. The current state of the art has been pushed forward by a combination of EFT and amplitudes techniques. The calculation of the waveform from amplitudes techniques
also benefits and exposes properties of the double copy.

Large scale structure surveys that provide the distribution of galaxies at lower redshifts are increasingly important windows into the universe. In many ways, the raw statistical power of large scale structure surveys rivals the CMB today and will rapidly exceed it with upcoming surveys. Unfortunately, our ability to use these maps to understand the universe is limited by our understanding of nonlinear structure formation and astrophysical uncertainties, not simply statistics. Theoretical insights paired with increasing sophisticated simulations have made many new and more powerful analyses possible, but much work remains if we are to harness the full power of these surveys for fundamental physics. On the largest scales, effective field theory and perturbation theory techniques
are sufficient to model both the DM, baryons and tracers thereof. Work on the effective theory
of large scale structure has made the regime of validity transparent, along with the systematic
expansion in the density, and has opened the door to analyses of the full-shape matter power
spectrum and bispectrum that have never previously been done. In addition to modeling the LSS perturbatively, theoretical understanding of both the signal and the non-linear evolution have yielded a number of protected observables, quantities that we can measure that are largely not impacted by nonlinear evolution, of which the baryon acoustic oscillations are one of the most significant examples.

On small scales, the density contrast becomes order one and perturbative methods break down. In addition, on these nonlinear scales baryonic feedback significantly alters both the distribution of DM and the formation of galaxies. DM-only simulations have been important in understanding this regime, but have left a number of outstanding problems concerning the distribution of matter on small scales, including the DM at cores of
galaxies and the number of satellites. Simulations of DM and baryons, including feedback from
active galactic nuclei and supernovae, have suggested that baryons could play a significant role in
addressing these problems. In contrast, a number of models of DM  have been proposed
that would alleviate these tensions between simulations and observations, by introducing small-
scale cutoffs in the matter power spectrum, giving DM self-interactions, or both. Continued
development of simulations on both cosmological and galactic scales are therefore essential to
our understanding of the fundamental physics of DM  and the astrophysics of galaxies and
galaxy clusters. 

\subsubsection{Inflation and acceleration - relation to fundamental theory}
 
The expansion of the universe is accelerating today and was very likely accelerating in the past,
due to dark energy and inflation respectively. These epochs present unique challenges for fundamental physics, both qualitatively and quantitatively. Recent progress has been driven by a
variety of advances connecting cosmology to the many corners of the theory frontier.  Yet,
understanding quantum gravity in cosmological spacetimes remains one of the largest and most
important unsolved problems in high energy physics, as it unites both basic theoretical questions,
cosmological observations, and even the origin and fate of the universe.

In recent years, our understanding of inflation has been significantly enhanced by the insights
of effective field theory, and the  EFT of Inflation has been developed. Inflationary model building makes extreme demands of the effective field theories that describe
the inflationary mechanism, which has inspired a wide range of work investigating the space of
consistent EFTs and their origins from a UV complete theory like string theory.  The search
for explicit models has yielded a variety of novel ideas that can be studied
and generalized in the context of inflationary EFTs, and yield new signals and analyses of
cosmological data.
In recent years, the constraints of self-consistency and the existence of a UV completion
have been applied directly to the EFT of inflation and the cosmological correlators. Often characterized as the cosmological bootstrap, this approach shares common elements with the
amplitudes and conformal bootstrap programs. One promising opportunity at the intersection of fundamental physics and cosmological observations is the impact of the particle spectrum during inflation, including light and heavy fields, on non-Gaussian cosmological correlators.

The discovery of the accelerated expansion of the universe today further challenges our under-
standing of fundamental physics. The small size of the cosmological constant has resisted a
natural explanation. In addition, evidence from string constructions and other theoretical
challenges have suggested that perhaps de Sitter space is fundamentally unstable and/or is one
of many accelerating regions. It remains entirely possible that our current acceleration is
more analogous to a second inflationary period and that de Sitter-like regions do not occur at
all. Observational constraints on this possibility are encoded in a similar effective field theory
description that is constrained by a variety of observational and gravitational probes. The challenges associated with de Sitter range from perturbation theory on a fixed de Sitter
background all the way to non-perturbative effects in a quantum theory of gravity. In recent years, significant progress has been made understanding the structure of perturbation theory in a fixed de Sitter background. Using EFT or diagrammatic techniques,
one can see both the IR and secular terms that arise in the cosmological slicing of dS can be resummed using the framework of Stochastic Inflation.

\subsection{Neutrino physics} \label{sec:nu}

The discovery of nonzero neutrino masses requires new fundamental fields and new interactions. We know very little about this new physics other than the fact that it exists. Among the goals of neutrino theory is to identify the different hypothetical degrees of freedom and interactions responsible for nonzero neutrino masses. More progress requires a coherent theoretical and phenomenological effort to establish connections to other outstanding questions in fundamental particle physics. 
On the phenomenology side, massive neutrinos must have nonzero dipole moments, and, since neutrinos and charged leptons are connected, neutrino mixing naturally leads to charged-lepton flavor violating processes. Being associated with neutrino masses, these quantities will likely receive contributions from the underlying new physics. Interpreting results from the corresponding experimental efforts in terms of the possible BSM theories requires robust theoretical predictions of the SM contributions to these processes.      
The coming precision era of neutrino oscillation experiments promises precise measurements of fundamental neutrino mass and mixing parameters, providing important information on the nature of these particles. In order to fully exploit the experimental measurements, precise theoretical predictions of the underlying neutrino-nucleus cross sections (including their energy dependence) are needed, which poses a significant theoretical challenge, as described below. 

\subsubsection{Neutrino masses and flavor models}

Within an effective theory approach (where the absence of any new light degrees of freedom beyond the SM is assumed) the neutrino masses originate from the unique dimension five operator of the form $-(LH)^2/(2\Lambda)$ with a suppression scale $\Lambda \sim 10^{15}$ GeV. In order to generate this operator one can safely conclude that there must be new degrees of freedom with masses at or below $10^{15}$ GeV. These could be either a neutral or a triplet fermion, or a triplet scalar, giving rise to various types of see-saw mechanisms. However, the mass of these new degrees of freedom could be anywhere in between the eV and the $10^{15}$ GeV scale.  Different new-physics scenarios leave imprints in different types of particle and nuclear physics probes and may be related to other outstanding questions in fundamental particle physics. A distinct possibility is that the neutrino mass is not captured by the effective operator, but instead there are new light degrees of freedom (right handed neutrinos) which form Dirac neutrinos together with the SM neutrinos, in which case a U(1)$_{B-L}$ symmetry is preserved. The Dirac neutrino scenario leads to the theoretical puzzles of why the relevant Yukawa coupling is many orders of magnitude smaller than even the electron Yukawa coupling, and how the exact global U(1)$_{B-L}$ symmetry remains unbroken by quantum gravity. 

The flavor symmetry breaking pattern of the lepton sector is quite different than for the quarks, and are often a more natural fit for discrete flavor symmetries, which circumvent some of the challenges faced by continuous symmetries. This approach to flavor also has its limitations including a certain level of arbitrariness when it comes to choosing how the symmetries are broken.
Flavor symmetries also provide concrete precision-targets for next-generation neutrino oscillation experiments. One example is the currently unknown CP-odd phase $\delta$ that parameterizes CP-violating effects in neutrino oscillations, and different flavor models make qualitatively different predictions for $\delta$. Consequences of flavor symmetries may also translate into relations among the different masses and mixing which might be tested experimentally.

\subsubsection{Neutrino phenomenology}

Neutrino oscillation experiments provide all affirmative information we have on neutrino masses. More, better neutrino oscillation data are expected in this and the next decade. Theory played a central role in the development of the formalism of neutrino oscillations, including the solution to the twentieth-century solar and atmospheric neutrino puzzles. This role is expected to persist and evolve as the oscillation probes grow more sophisticated and diverse.  Different neutrino-phenomenology research groups provide the best-fit values for the different oscillation parameters that are used by the particle physics community as a whole. New neutrino oscillation data can also reveal more new physics in the neutrino sector, including  new weaker-than-weak
neutrino–matter interactions that modify neutrino production, detection, and flavor-evolution, new light neutral fermions that lead to the presence of new oscillation frequencies or to the apparent non-unitarity of the leptonic mixing matrix, or allowing for the possibility that neutrinos have a finite lifetime.  

Neutrino theory also aims at interpreting unexpected results in neutrino experiments collectively referred to as the short-baseline anomalies. While there is currently no outstanding solution to the short-baseline anomalies, it is possible that the correct answer
is still to be uncovered, that may include the existence of new physics. These could include, for example, new neutrino states that participate in new interactions. Since neutrinos interact only via the weak interactions, neutrino scattering experiments are especially sensitive to new, weaker-than-weak interactions mediated by hypothetical light or heavy new particles. 
Neutrino scattering experiments (which are fixed-target spanning an energy range of more than 20 orders of magnitude) can reveal the existence of new interactions, new neutrino properties, or new particles.

 As explained, neutrino masses imply the existence of new particles. If these are light enough
and strongly-coupled enough, they can be produced and detected at high energy colliders. For example right-handed neutrinos with Majorana masses of order tens to thousands of GeV
could  be produced at the LHC and other high energy colliders, just like ordinary neutrinos.  Neutrinos and charged-leptons are intimately connected. The fact that neutrinos mix, for example, implies that lepton-flavor numbers are not conserved in nature so flavor-violating processes involving charged-leptons are also guaranteed to occur. In the absence of new interactions, however, the expected rates for charged-lepton flavor violating (CLFV) process, including $\mu\to e\gamma$, $\mu\to 3e$, and $\mu\to e$ conversion in nuclei, are tiny. New physics, including the interactions and degrees of freedom that may be responsible for nonzero neutrino masses, can lead to much
larger rates for CLFV, often providing the best constraints on some models. Neutrino oscillation experiments often also allow other particle physics measurements and searches for new phenomena, many of which are only peripherally related to neutrinos. For example, near detectors in neutrino oscillation experiments are an excellent place to search for light, hidden sector particles,
including a dark photon produced in neutral pion decay, while far detectors are ideal for monitoring a large number of nucleons in search for proton decay and other baryon-number violating processes.

\subsubsection{Neutrinos in astrophysics and cosmology}

The past decade has seen important advances in neutrino astrophysics and cosmology. These advances inform us about the nature of the sources from which the neutrinos originate, but also inform us on the properties of neutrinos. High energy neutrinos (from GeV and above) range from atmospheric to GZK neutrinos, and are usually correlated with cosmic rays.  IceCube provides very high statistics atmospheric data at energies from a few GeV
to a few TeV, which can be used to reduce some of the
systematic uncertainties associated with atmospheric neutrinos, also opening up
the possibility of measuring a larger subset of oscillation channels, including $\nu_\mu \to \nu_\tau$. Over the last decade IceCube also observed neutrinos with energies well above a TeV coming from outside our
galaxy, which provide a unique avenue to probe BSM effects, given the very long distances traveled by
the neutrinos, as well as their ultra-high energies. 

The detection of low-energy astrophysical neutrinos is an experimental and theoretical challenge onto itself;
theory and phenomenology is required in order to understand how these neutrinos “look” inside different
detectors.  Supernova neutrinos provide a  unique laboratory for neutrino physics and astrophysics, since apart from the Big Bang supernovae are the only environment in the universe where neutrinos are in or near thermal equilibrium
conditions. The detection of neutrinos from the next individual supernova contains invaluable information on neutrino
properties. However extracting information will require understanding the evolution of neutrino flavor inside the supernova explosion. Solar neutrinos already found spectacular application by pinpointing the LMA-MSW solution to neutrino flavor transformation from the Sun to the Earth, nevertheless still there are outstanding questions that surround some of the data.

Cosmological neutrinos produced in a thermal spectrum shortly after the Big Bang contain a wealth of information on particle physics, astrophysics and cosmology. Neutrinos contribute a significant component of the energy density in the early universe, which can be observed through
precision measurements of the CMB and through other probes of the recombination of the universe.  Neutrinos also affect the growth of structure through a change in the background expansion rate of the universe,
and through their significant thermal velocities. Upcoming
cosmic surveys such as CMB-S4, Euclid and the Vera Rubin Observatory will be sensitive to the sum of the neutrino
masses. Extracting neutrino masses from these cosmic surveys is a nontrivial exercise that requires dedicated theoretical analyses and computations, involving for example, many-body physics and nonlinear dynamics.

With their vastly disparate energy scales, observations of astrophysical and cosmological neutrinos provide a wealth of information that is complementary to terrestrial experiments, entwining astrophysics, cosmology, and particle physics phenomenology. Theory and phenomenology play a key role in using this information to deepen our understanding of neutrino properties and interactions. In addition, theoretical work at this interface may yield insights into the properties of astrophysical objects and explosive phenomena and of the medium the neutrinos traverse.

\subsubsection{Neutrino cross sections}

Maximizing the discovery potential of neutrino experiments in the coming precision era requires theoretical predictions for the underlying neutrino-nucleus cross sections with commensurate (or better) precision to experiment. The large range of neutrino energies makes this a complicated task, requiring a coordinated effort from phenomenologists, lattice field theorists, nuclear theorists, and computational theorists working on event generators.  
To understand the energy dependence of the cross sections, separate theory inputs for the processes which dominate at the different energy scales are needed. The range covers coherent and inelastic scattering at $\lesssim 100$~MeV, quasi-elastic scattering (QE) dominated by single-nucleon knockout processes at $0.1-1$~GeV, two-nucleon emission and resonant production at $1-3$~GeV, shallow inelastic scattering (pion production) at $\gtrsim 3-5$~GeV, and deep inelastic scattering (DIS), where the neutrino can resolve the individual quark constituents of the nucleon at $\gtrsim 5$~GeV. Each of these regimes requires knowledge of both the nuclear ground state and the electroweak coupling and propagation of the struck nucleons, hadrons, or partons through the nucleon. The range of challenges is extreme; QE scattering and DIS are conceptually the easiest to understand, but also require nonperturbative inputs to quantify. In order to obtain precise theoretical cross section predictions with fully quantified uncertainties, a robust pipeline is required combining inputs from perturbative and nonperturbative QCD and nuclear EFT with event generators.  Ultimately, precise predictions of inclusive and exclusive cross sections across a wide range of kinematics are required for the success of the precision era of neutrino experiments.

Bridging the gap from precise theoretical predictions to experimental data requires the development of sophisticated event generator tools. The current- and next-generation neutrino experiments are reaching a point in which statistics are no longer the dominant uncertainty. The largest systematic uncertainty is the modeling of the primary interactions and the propagation of hadrons from the primary interaction out of the nucleus. Improving this falls under the purview of event generators. Further development of neutrino event generators is essential to meet the needs of DUNE and the SBND program. Additionally, with the general purpose near detector of DUNE, it will be vital to develop the ability to efficiently investigate BSM processes through automation developments within generators, similar to how this is handled at the LHC. Finally, there are many open questions that can be best answered by theorists involved in event generation, such as the modeling of hadronization at $Q^2 < 10  $ GeV$^2$ and the matching from shallow inelastic scattering to DIS.

The theoretical description of the large nuclei employed as targets in neutrino scattering experiments is based on nuclear many-body effective theory and requires nonperturbative inputs for single- and few-nucleon form factors and matrix elements. While electron scattering experiments provide precise inputs for vector current elastic form factors, axial current elastic form factors are not well known. Dedicated efforts already exist to calculate the needed form factors in lattice QCD with good control over systematic errors. Lattice-QCD based nucleon form factors and quasi-elastic scattering inputs for nuclear many-body theory with total uncertainties at the few-percent level are within reach of current theoretical methods and available computing resources (see Sec.~\ref{sec:lft}). 

Resonant and non-resonant nucleon axial transition form factors, needed at higher neutrino energies, are also accessible from lattice QCD. While first results are already available, more effort is required to deliver results with complete error budgets. Nuclear many-body theory predictions on nuclear modifications to resonant and non-resonant pion production and absorption processes are also needed. Lattice-QCD studies of PDFs, relevant at yet higher neutrino energies in the DIS region, are rapidly maturing. The transition regions between low- and high-energy theories use different degrees of freedom to describe neutrino-nucleus interactions which poses an important theoretical challenge requiring further studies. 

\subsubsection{Neutrinoless double beta decay}

The next generation of tonne-scale neutrinoless double-beta ($0\nu\beta\beta$) decay experiments has the opportunity to answer fundamental questions about the nature of neutrino masses: observation of $0\nu\beta\beta$ decay will imply that neutrinos are Majorana particle. In addition these experiments are sensitive to a variety of lepton-number-violating (LNV) mechanisms, including the standard scenario driven by the exchange of light Majorana neutrinos, low-scale seesaw scenarios
with light sterile neutrinos below the electroweak scale, and models of physics beyond the Standard Model (BSM) with new degrees of freedom at the TeV scale. The interpretation of  $0\nu\beta\beta$ experiments demand an ambitious theoretical program including further developing particle-physics models of LNV,  and  computing $0\nu\beta\beta$ rates with minimal
model dependence and quantifiable theoretical uncertainties.

\section{Computational theory and Quantum Information} \label{sec:comth}

Computational theory, the third Theory Frontier theme, seeks to quantify predictions of the underlying QFT in regions beyond the reach of semi-analytic methods. As such it typically starts from a well-defined underlying theoretical structure which can be evaluated using computational methods. 

All areas of high energy theory benefit from or contribute to the development of computational methods to some extent. In collider phenomenology and precision phenomenology based on multi-loop perturbation theory, computational methods are crucial in all components that are needed to obtain reliable theoretical predictions. A brief overview of the challenges surrounding event generators, which connect theoretical calculations to experimental observations, is given in Sec.~\ref{sec:coll}. The multi-loop amplitudes at the core of the perturbative calculations contain complex mathematical structures and theoretical structures requiring specialized computational strategies. The integration over final state momenta to obtain the hard, parton-level cross sections is further complicated by the presence of infrared and collinear singularities. Finally, the PDFs are nonperturbative inputs, currently obtained from fits to large experimental data sets, presenting their own set of computational challenges. Cosmological, astrophysical, and DM simulations are central to theoretical research in astro-particle physics and cosmology. The development and refinement of computational methods for these simulations is crucial to  progress in this area, as discussed in Sec.~\ref{sec:cos}. 

A classic computational theory example is Lattice Field Theory (LFT). The standard approach to LFT starts from the Euclidean (imaginary-time) path integral to study  nonperturbative effects of strongly coupled theories of interest, including (but not limited to) QCD, in Monte Carlo simulations on classical computers. Quantum Computing and Quantum Simulation of QFTs hold great promise to overcome limitations of Euclidean LFT for the study of systems such as real-time scattering dynamics, finite density systems (quark gluon plasma, neutron stars), and QFTs with $\theta$ terms, among others.  

New theoretical methods and insights are also important drivers of progress in computational theory, while computational demands can provide motivation for theoretical work. For example, insights gained from the amplitude program developed in fundamental theory have enabled progress towards N$^3$LO calculations in multi-loop collider phenomenology. On the other hand, the complex analytical structures of the multi-loop amplitudes and partonic cross sections invite development of novel EFTs and the study of special mathematical functions. More recently, amplitude methods are employed to obtain gravitational wave templates. To list just one example for lattice QCD, the connection between discrete finite volume spectra of multi-hadron systems to scattering phase shifts, discovered long ago by L{\"u}scher, has been developed into a powerful formalism to obtain scattering amplitudes, resonance properties, and weak decay amplitudes into resonances, however limited at present to systems with at most three hadrons. A new theoretical method relates finite-volume matrix elements of current-current operators to inclusive scattering or decay rates, which could open up an entirely new set of observables to lattice-QCD studies in the coming decade.     
Connections between fundamental theory and computational theory are discussed in Sec.~\ref{sec:qft} for the numerical conformal bootstrap. While insights from fundamental theory can potentially drive new computational methods, computational theory, in turn, can provide powerful tests of analytic predictions of the properties or dynamics of QFTs (or other theoretical structures in question), potentially yielding a deeper understanding of the structure of the underlying QFT. 

Computational theory is tied to enabling technologies, both driving and benefiting from ever more sophisticated and powerful platforms for classical computers, quantum computers and quantum simulators, as well as algorithm development and code optimization.  In the development of LFT as a general purpose, precision tool over the past 40 years, interdisciplinary efforts connecting to computer science and applied mathematics were important drivers of progress. A famous example is the co-design of the IBM Blue Gene series in the 2000s. Recent examples involving Machine Learning and Artifical Intelligence (ML/AI) are discussed below.  

\subsubsection{Machine Learning and Artificial Intelligence}

Machine Learning and Artificial Intelligence describe a broad class of learning algorithms, including for example, deep learning architectures based on complex neural networks with many layers. ML methods have been developed for and used in high energy experimental data analyses for many years. ML is now playing an increasingly important role in all areas of high energy theory, in particular, those that rely on theoretical simulations or numerical analysis. 
In particular, novel ML methods are being developed or adapted in first-principles simulations in lattice field theory and event generation. In LFT, ML applications have been developed for all stages of the computations, including the generation of gauge fields and correlation functions as well as the numerical analysis needed to extract the physical observables of interest from the underlying correlation functions. Similarly, in standard event generators, all modules can be improved through ML, including phase space sampling, scattering amplitudes, loop integrals, parton showers, parton densities, and fragmentation. End-to-end ML generators, which use generative  networks to directly generate parton-level events, complement standard generators in important ways. 
The generative ML models that have been developed for efficient sampling of gauge fields in LFT or phase space in event generators are based on new symmetry-preserving ML architectures that can include correction steps to guarantee exactness. This development is a nice example of cross-disciplinary synergy, as these new networks are finding applications in other fields of science and also in industry, inviting further cross-disciplinary collaboration. 

Other examples, include jet structure in collider phenomenology, where, as discussed in Sec.~\ref{sec:coll}, ML is used to construct new data-driven observables building on synergy between theory and experiment. 
For BSM phenomenology studies, generative ML and other simulation-based inference frameworks have also proven useful in efficiently sampling  high-dimensional BSM parameter spaces (e.g.\ supersymmetric theories) to check for consistency with experimental data. In cosmology, ML networks can be used to accelerate hydrondynamics simulations, which can, in turn be used in simulation-based inference frameworks to constrain cosmology and galaxy formation parameters. 

\subsection{Lattice Field Theory (LFT)} \label{sec:lft}

The development of computational methods for the study of lattice field theory was originally motivated by the desire to understand QCD in the nonperturbative regime. Since then, lattice QCD has been developed into a precision tool with applications to a broad range of observables. However, LFT is also used to study other QFTs with interesting, strongly-coupled dynamics, as we describe in the next subsection.  

\subsubsection{Lattice QCD}

Lattice QCD is strongly connected to Phenomenology (the second Theory Frontier theme), as nonperturbative effects from QCD are ubiquitious wherever quarks or hadrons are involved. Thanks to remarkable progress in the past decade, lattice QCD provides essential inputs for quantities in all areas of Phenomenology discussed in Sec.~\ref{sec:pheno}. In the coming decade, the list of quantities for which lattice-QCD inputs meet the precision goals of current and planned HEP experiments is expected to grow significantly, a result of theoretical, computational and technological advances as well as continuing dedicated efforts. 

In particular, lattice QCD plays a crucial role in interpreting experimental measurements in flavor physics, where physical processes involve hadrons in the initial or final states, or where the effects of virtual hadrons are important. Following decades-long, sustained efforts in the US (and internationally), precise lattice-QCD results are now available for a large number of hadronic matrix elements  relevant for weak decays of light, strange-, charm-, and beauty-flavored hadrons, in all cases with completely quantified error budgets and in some cases with commensurate (or better) precision compared to experiment, hence maximizing the discovery potential of the experimental program. With one notable exception in the kaon system discussed below, the majority of results concern ``simple quantities'', defined as observables involving local operators with at most one hadron in the initial and final states respectively, which are (almost) stable under the strong interactions. 
These lattice-QCD results can either be used in determinations of fundamental parameters of the SM (when combined with corresponding experimental measurements) or to confront SM theory with experiment. Indeed, subpercent calculations of light-meson decay constants and semileptonic form factors have enabled precise determinations of CKM elements from the corresponding experimental measurements. Subsequent unitarity tests for the first row of the CKM matrix have recently yielded a surprising tension, prompting further investigations. Lattice-QCD results also enable precise SM predictions for rare decay processes, and improve SM predictions for the LFU testing ratios discussed in Sec.~\ref{Sec:Precision}. Here the comparison between SM theory and experiment provides stringent constraints on BSM theories or could yield discovery of new physics, should a significant ($>5\sigma$) disagreement between theory and experiment be observed. Within the next five years, (sub-)percent-level lattice-QCD results will become available for the whole suite of semileptonic $B$, $B_s$, $D$, $D_s$ form factors and mixing matrix elements, given continued, sustained effort.
Similar calculations of baryon form factors, while less precise, are also feasible with current methods. These results will complement the expected experimental measurements at Belle II, LHCb and BESIII, enhancing the discovery potential of these experiments. They may also shed light on the observed tensions between determinations of the CKM parameters $|V_{cb}|$ and $|V_{ub}|$ from exclusive vs.\ inclusive semileptonic $B$-meson decays, respectively. Lattice-QCD calculations of inclusive $B$-meson decay rates could provide more direct probes of the exclusive-inclusive tensions. New theoretical methods for such computations are under active development and may provide a path towards such results.   

The muon's anomalous magnetic moment ($a_\mu$), with a tantalizing $4.2\sigma$ tension between SM theory and experiment, is a prominent example of the discovery potential of precision physics. The anomaly arises from contributions of virtual particles, where all SM particles participate. The uncertainty in the SM prediction is driven by low-energy hadronic contributions, which are difficult to quantify precisely due their nonperturbative nature. Historically, in lieu of direct calculations,  dispersion theory combined with data measurements of $\sigma(e^+e^- \to {\rm hadrons})$ (so-called data-driven evaluations) were used to quantify the dominant hadronic vacuum polarization (HVP) contribution to $a_\mu$. Indeed, the significance mentioned above is obtained from dispersive evaluations of HVP based on current, precise cross section measurements. While most existing lattice-QCD results for HVP are too uncertain for a meaningful comparison, a first lattice-QCD calculation with sub-percent precision is now available, albeit in tension with the data-driven prediction implying a reduced significance with the experimental average. This surprising result is currently being scrutinized in independent lattice-QCD calculations with commensurate (or better) precision. Indeed, given the dedicated efforts by the lattice community, it is expected that lattice-QCD results for HVP will meet the target precision set by the Fermilab experiment, provided that the different lattice-QCD results agree with each other. Here the comparison between lattice-QCD and data-driven HVP results will be crucial to the interpretation of the experimental measurement of $a_\mu$. 

A growing number of lattice-QCD calculations of increasingly complex quantities with control over systematic errors are becoming available. Prominent examples are results for the $\Delta I = 1/2$ rule in nonleptonic kaon decay and the direct CP-violation parameter $\epsilon'/\epsilon$ in the kaon system. They were made possible, in part, by theoretical developments, including relating finite volume matrix elements to decay amplitudes and the formulation of chiral fermions. There are many other quantities for which sufficiently precise lattice-QCD calculations are needed to interpret quark- and lepton-flavor physics experiments.
These needs are being addressed in a broad program of lattice-QCD calculations, which includes development work to establish the needed precision or expand methodologies to increase the scope of the calculations. Ongoing work includes projects to compute semileptonic form factors for decays of heavy-flavored hadrons into vector-meson final states, isospin-breaking corrections (including radiative QED effects) in simple weak decay matrix elements, and rare kaon decay amplitudes, among others.

Precision tests of the SM nature of the Higgs boson confront measurements of Higgs-boson branching fractions with SM expectations, which are affected by parametric uncertainties from the quark masses and $\alpha_s$. Lattice-QCD based determinations of these fundamental parameters already meet the precision demands of the LHC program in the coming decade. Theoretical predictions of, for example, LHC cross sections or neutrino-nucleus cross sections in the deep inelastic regime require $x$-dependent PDFs as inputs. These nonperturbative functions are related to light-cone correlations of quark and gluon fields inside a given hadron, which are difficult to compute directly in a Euclidean framework, such as lattice QCD. The past decade has seen the development of several promising approaches and intense efforts to implement them in numerical computations and study their relative merits. They include quasi-PDFs which are related to the usual PDFs via a new EFT, ``Large Momentum EFT''; pseudo-PDFs, related to Ioffe-time distributions; and Compton amplitude and current-current matrix elements coupled with OPE and/or inverse transforms, for which new strategies are under active development as well. In addition, other multi-variable matrix elements, such as GPDs (Generalized Parton Distributions) and TMDs (Transverse Momentum Distributions) are also being studied. There are various technical, computational, and theoretical challenges associated with each approach, and ongoing efforts seek to obtain lattice-QCD results with control over the dominant systematic effects. Once such results are available, they can be tested against PDFs obtained from experimental data. Lattice-QCD results for PDFs will be particularly important in regions that are less-well determined from experimental data. 

Nucleons and nuclei are commonly used in searches for violations of fundamental symmetries of the SM and for BSM signatures. Thanks to growing, dedicated efforts, lattice-QCD results with few-percent-level uncertainties have recently become available for a key nucleon quantity (the iso-vector axial charge, $g_A$). This can be regarded as marking the start of the precision era for lattice-QCD calculations in the nucleon sector. Assuming continued effort, coupled with increases in computational resources, and leveraging new computational methods to address the signal-to-noise problem in nucleon and nuclear correlation functions, we expect that the coming decade will see a growing number of lattice-QCD results for nucleon matrix matrix elements with completely quantified uncertainties, possibly reaching, in some cases, percent-level uncertainties. Among these quantities are the nucleon vector, axial-vector, and pseudoscalar form factors, which are needed as inputs for theoretical predictions of neutrino-nucleus cross sections as already discussed in Sec.~\ref{sec:nu}; nucleon EDM matrix elements of SM and BSM operators, relevant for planned neutron and proton EDM experiments; the scalar, axial, vector, tensor charges of the nucleon, which are important to interpret high-precision neutron-decay experiments yielding constraints on CP violation and BSM interactions; and spectroscopic studies of excited state hadrons and resonances relevant for precision EW hadronic decay processes and for neutrino-nucleus cross sections. Nucleon matrix elements describing proton decay, neutron-antineutron oscillations, muon to electron conversion, and neutrinoless double beta decay are examples of other quantities where lattice-QCD inputs are critically needed for theoretical prediction of the corresponding rates. Based on ongoing efforts, it is entirely reasonable to expect that lattice-QCD results with fully quantified uncertainties will become available in the coming decade for at least some of these other quantities. 

Given the examples above, it is clear that theoretical calculations of a wide range of nucleon and nuclear properties, which are also of interest in the context of nuclear physics, are essential inputs for important, low-energy particle-physics experiments. These experiments often use heavy nuclear isotopes, e.g., Argon at DUNE and Germanium in certain neutrinoless double-beta decay and dark-matter detection experiments. In order to obtain quantitative information about the underlying short-distance (B)SM physics, the nuclear responses to the EW probes must be quantified with sufficient precision. An {\em ab-initio} QCD description of these complex nuclear systems is well out-of-reach of current lattice-QCD methods. More suitable theoretical treatments account for the vast range of scales involved, and use nuclear many-body theory and nuclear effective theory to capture the very low-energy nuclear effects. The role of lattice QCD then is to provide the non-perturbative inputs for matrix elements of the various SM and BSM operators in single- and few-nucleon systems to constrain the corresponding EFTs. These EFTs can then be used in many-body nuclear-structure calculations to evaluate the relevant observables in heavier isotopes. This program requires a coordinated effort between the  high-energy and nuclear theory communities. First lattice-QCD results for  matrix elements of two-nucleon systems as well as of light nuclei ($A\leq4$) already exist, albeit with unphysically heavy pion masses and other limitations. Further efforts to develop more efficient computational strategies as well as significantly larger computational resources are required in order to obtain lattice-QCD results for such systems with full control over the statistical and systematic uncertainties.

\subsubsection{Lattice Field Theory beyond QCD}

LFT provides access to the nonperturbative properties of asymptotically free Yang-Mills theories, where the gauge group (typically SU($N_c$), but not limited to it), the number of fermions ($N_f$) and their properties (masses, representations) can be adjusted depending on the case at hand, among other variations. This allows for the study of specific classes of theories to obtain general nonperturbative statements about these theories, for example, studying the $N_c$ dependence in the 't Hooft and/or Veneziano limits. It also allows for LFT studies of specific theories, to investigate, for example, the nonperturbative properties of a given BSM theory. 

Conformal field theories (CFT) are broadly interesting in high energy and condensed matter theory. The study of emergent conformal symmetry in Yang-Mills theories coupled to many fermions (\ie,  SU($N_c$) gauge theories with $N_f \gg 1$ fermions in fundamental or higher representations) is an interesting target of LFT investigations. While identifying the location and order of the expected conformal transition is an ongoing challenge, these studies have yielded phenomenologically and theoretically interesting discoveries. In particular, the existence of a light scalar bound state, which is a possible candidate for a pseudo-dilaton, has been established by several groups, prompting in turn the formulation of dilaton EFT (where such a state would emerge near the conformal transition). These results invite additional nonperturbative studies to further develop the dilaton EFT description. They are of phenomenological interest as potential candidate theories for composite Higgs models, where LHC constraints could be satisfied if the composite Higgs is a pseudo-dilaton or pseudo Goldstone boson. 
With the development of general purpose tools such as gradient-flow RG, further studies to locate the conformal transition and obtain information about the properties of the emergent CFTs may be feasible yielding results potentially relevant to numerical conformal bootstrap, for example. 
LFT studies of these and related theories also provide interesting opportunities for BSM phenomenology. LFT results for low energy constants combined with experimental bounds may provide constraints on BSM theory space. Conversely, if strongly favored UV completions are identified, their phenomenology may be obtained from LFT computations. 

Composite dark matter models, where the dark sector is described by a (QCD-like) confining theory, are natural targets for LFT studies. Here the methods developed for lattice QCD can be straightforwardly applied to obtain information about the thermodynamics, spectrum, and form factors of the DM particles. The thermodynamics, in particular the nature of the finite-temperature transition of the DM model, affects early-universe cosmology, with the intriguing possibility that a first-order transition could leave an imprint of primordial gravitational waves. Self-interactions can also be studied with the same methods as developed for hadron-scattering in lattice QCD.  Addressing questions such as the dark-matter relic abundance implied by the DM model, while out of reach with current methods, may become feasible with methodological advances. 

In summary, studies of relevant LFTs are highly desirable in order to more fully explore and constrain BSM theory space with quantitative information on strongly-coupled QFTs. Indeed, as outlined above, LFT studies of select theories have already yielded very promising results. Interestingly, existing lattice-QCD results obtained at unphysically large up/down-quark masses can be reused in Neutral Naturalness scenarios, where a broken mirror symmetry leads to a QCD-like hidden sector with heavy mirror-quarks. Other potential targets include theories of right-handed neutrinos or composite axions. The large space of theories amenable to LFT computations provides opportunities for exploration, hindered, however, by the high computational cost. Here, theoretical innovations in the theory of ML to efficiently sample configuration space may have transformational impact to alleviate the computational cost. This could enable timely studies of a broad range of relevant LFTs in the coming decade. 

Supersymmetry, even if not realized in Nature, is an essential ingredient in the AdS/CFT correspondence and studies of supersymmetric QFTs continue to yield valuable theoretical insights (see Sec.~\ref{sec:qft}). A recent notable development is a novel lattice formulation of ${\cal N} = 4$ supersymmetric Yang-Mills (SYM) theories which preserves a subset of the full continuum supersymmetry to avoid fine-tuning in the continuum limit. First results from lattice studies in this formulation find good agreement with holography at surprisingly low values of $N_c$. This opens the door to a whole new area of study with strong synergy with semi-analytic treatments. 

\subsection{Quantum Information} \label{sec:qis}

Quantum information has played an increasingly important role in high energy theory research in recent years. Work on the theory of quantum information has yielded new insights into quantum gravity and revealed new structures in certain QFTs. Methods are being developed for quantum computations of QFTs relevant to HEP, offering the intriguing possibility of access to nonperturbative regimes in systems out of reach of classical computation and semi-analytic methods. As discussed in Sec.~\ref{sec:cos}, theorists are deeply involved in leveraging quantum sensor technology to enable novel experiments.   

\subsubsection{Quantum simulation and quantum computing}

Motivated by the limitations of classical computation, recent years have seen a dedicated effort to develop the theoretical formulations, methods, and algorithms to simulate QFT systems on quantum computers as well as analog quantum simulators. The hope is that, as quantum hardware and computational methods mature, it will be possible to employ quantum computations to obtain information about problems such as real-time dynamics of hadron collisions, early universe evolution, the neutron-star equation of state, the nature of QCD phases (including the quark-gluon plasma), SM and BSM theories with chiral fermions or $\theta$ terms, axion cosmology, and other systems with sign problems. 

The Hamiltonian formulation is a natural choice for quantum simulations, where it is possible to represent the dynamics such that the number of required building blocks (qubits and gates) depend only polynomially on the number of the degrees of freedom. However, the Hilbert space of quantum field theories is infinite-dimensional and must still be truncated in some fashion. In classical lattice field theory simulations space-time is discretized, while in quantum simulations truncation in field space is also necessary. 
A number of different digitization schemes have been proposed and are being studied, including Casimir dynamics, conformal truncation, discrete groups, dual variables, light-front quantization, loop-string-hadron formulation, quantum link models, and qubit regularization, among others. The effects of the truncation or digitization schemes need to be carefully studied, particularly for gauge theories, where truncations that break gauge invariance can, in principle, render the theory inconsistent. As a general strategy, truncations that preserve more of the symmetry are favored, as are local encodings of the qubits.  
After choosing a truncation scheme and encoding the degrees of freedom onto the quantum hardware, the next step is the initial state preparation, formulated in terms of the fields of the theory, followed by time evolution. In the case of digital quantum hardware, the time evolution operator needs to be approximated, for example, via ``trotterization'', which truncates the infinite series representing the time evolution operator. 
Efficient quantum algorithms with tightly bounded errors on observables and concrete quantum-computational resource requirements need to be developed for all the relevant problems in high-energy physics, an endeavor that has started in recent years and needs extensive theory input.

In the case of analog quantum simulations, the physical system (for example, cold atoms) and its set-up to control the quantum states are designed to mimic the desired quantum field theory. Each quantum simulator hence represents a dedicated experiment, where the ``quantum simulation'' requires an initial state preparation and measurements after time evolution. While the space of possible target theories may be limited with a given experimental set-up, analog simulators may be attractive as the experiments can be scaled up to accommodate significantly larger systems which can describe larger Hilbert spaces compared to present-day digital quantum computers. 

Tensor network methods, which originated in condensed matter physics, can be used to reformulate lattice gauge theories into  fully discrete formulations suitable for Hamiltonian simulation and quantum computation.  In one approach, local tensors provide translational invariant building blocks of exact discretizations of the path integral, and encode both the local and global symmetries of the original theory. Symmetry-preserving truncations can then provide controllable finite-dimensional approximations with the same continuum limit as the original theory.  In another approach, tensor networks can approximate the wavefunction of a quantum many-body system efficiently, where the dimensionality of the tensors (namely the bond dimension), represents the amount of entanglement retained in the construction of the wavefunction. Hamiltonian dynamics can then be implemented by expressing the operators acting on this tensorial space. Progress in recent years is focused on pushing the developments to higher dimensional and non-Abelian lattice gauge theories. Tensor network constructions can be used to perform classical simulations of quantum circuits, which would yield valuable benchmarks for quantum simulations, crucial tests of the results, and inform aspects such as state preparation.

Given current NISQ (noisy intermediate-scale quantum) era limitations of quantum hardware, theoretical efforts are focused on low-dimensional systems where a variety of different QFTs, including U(1) and SU(2) gauge theories are being studied. Current efforts to understand the best truncation schemes for gauge theories will be very valuable in the future as more powerful quantum computers are being developed.
Important tasks for the near-term future include developing the building blocks and optimizing the approximations for NISQ machines. Other potential near-term applications include optimizations for classical computations, for example, of interpolating operator constructions in LFT calculations. 
The insights gleaned from developing quantum computing methods for QFTs of interest to HEP will also benefit quantum computing in other fields.

\subsubsection{Fundamental aspects of QIS}

The past decade has seen a remarkable convergence of quantum gravity and quantum information, as detailed earlier in this report (see Sec.~\ref{sec:funth}). Moreover, quantum information provides a new perspective on quantum field theory.  In the consideration of fundamental aspects of quantum information theory, entropy, entanglement, quantum error correction and quantum communication play a prominent role. Unlike traditional explorations of QFTs, information content is used as the organizing principle to better account for the  underlying quantum nature of the systems. This has led to the recognition that the structure of entanglement plays a key role in the properties of QFT systems. Quantifying the amount of entanglement to characterize the quantum information depends on the entropic function. Using entanglement entropy in the holographic context reveals a connection between entanglement and geometry, a profound result. 

Also significant are the connections that have been revealed between quantum information and nonperturbative structures of QFTs, including irreversability of renormalization group flows, classification of topological phases, implications for dynamics, and progress in understanding the black hole information paradox using holographic entanglement. Information theoretical methods have been used to prove so-called null energy conditions, which place tight constraints on the QFTs. The spread of quantum information in QFTs and other complex systems obeys universal laws and is closely related to transport properties. The study of out-of-time-order correlation functions has yielded insights into scrambling and bounds on the onset of quantum chaos. In short, information theoretic methods applied to QFTs have already yielded important insights, leaving the field wide open for further explorations along themes such as the connections between gauge invariance and information, towards seeking a deeper understanding of the nature of quantum field theory and quantum gravity. 

\section{Community Development}

The theory community in the United States is world-leading, a remarkable achievement given the extraordinary level of its international partners. Continuing excellence in research is necessary for maintaining this leadership, but not sufficient; robust support and community development is needed. The United States should emphatically support a broad and balanced program of theoretical research covering the entirety of high-energy physics, from fundamental to phenomenological to computational topics, both in connection to experiment and in its own right. Such support should facilitate the exploration of new research directions, the incorporation of new developments from adjacent fields, and the building of new bridges to gravity, cosmology, astrophysics, quantum information, nuclear physics, AMO, condensed matter physics, statistics, computer science, data science, and mathematics. 

The theory community is most effective as part of a balanced HEP program of Projects and Research, as both are essential to the health of the field.  Within this balance, support for {\it people} is vitally important, as they constitute the primary infrastructure of Research endeavors. As such, sustaining the vibrancy of the US theory community requires a concerted effort to support the people that comprise it (especially at early career stages), strengthen the fabric of the community, and broaden participation. Theory contributes an important cultural element within university departments and the broader physics community. Theorists initiate students into the frontiers of particle physics and lead them to fluency in quantum mechanics and field theory, while providing intellectually stimulating environments at their respective institutions. A key goal is to maintain a program across HEP that trains students and junior scientists, providing them with continuing physics opportunities that empower them to contribute to science. 

The success of our endeavors depends on ``$4 \pi$'' coverage in identifying and cultivating talent at all career stages.  To this end, the US theory community (as part of the broader HEP enterprise) would benefit from support to enable the realization of many of the recommendations made in the CEF02 (Career Pipeline and Development in Particle Physics) and CEF03 (Diversity, Equity, and Inclusion in Particle Physics) topical group reports. These include e.g.~strengthening support for early career physicists; emphasizing workforce development and reinforcing career pathways in both academia and industry; bringing awareness to the community about different forms of marginalization; creating pathways into the field for members from marginalized backgrounds and providing the support necessary for their success; engaging communities from emerging and developing countries; involving outside experts to help develop strategies for improvement; improving awareness and support for mental health issues; and cultivating work-life balance.

The effectiveness of the theory community as a part of the broader HEP enterprise is enhanced by targeted bridge-building initiatives that connect theory to experiment or to enabling technologies or to both. Past and present examples of such highly-effective initiatives include the LHC Theory Initiative, the LQCD Project, the Neutrino Theory Network, the Muon $g-2$ Theory Initiative, the HEP-QIS QuantISED program, AI for HEP, the Exascale Computing Project, and SciDAC. Support for ongoing and emerging initiatives will strengthen connections to experiment and sharpen focus on key HEP scientific objectives.

\section{Conclusion}

As we enter an era with many promising experiments but few guarantees of discovery, theory is as important as ever. Theory unifies the frontiers of particle physics. It is essential to the conception, execution, and interpretation of current experiments. Theory is an essential driver of the development and implementation of new, enabling technologies. It also extends well beyond by laying the foundations for future experiments and advancing our understanding of Nature in regimes that experiment has yet to reach. A robust theory program is vital to the success of current and future Projects in particle physics.

As summarized in this report, the past decade has seen remarkable (and deeply interconnected) progress across all facets of high-energy theory, from fundamental theory to phenomenology to computational theory. This progress --- alongside rapid developments at the accelerator and the computational frontiers; data from myriad experiments at the cosmic, energy, neutrino, and rare and precision frontiers; and convergence with other fields of physics  --- has contributed greatly to the advancement of particle physics and laid the groundwork for a coming decade of immense promise. 

The theory community in the United States has been central to the recent developments discussed in this report and is poised to play a leading role in the years to come. Such leadership is invaluable and cannot be taken for granted. We do not know what Nature has in store for us. We will be in the best position to realize the immense promise of the coming decade with strong and sustained support for all aspects of theory, both in relation to Projects and in its own right.

\section*{Acknowledgements}

We are grateful to the many members of the high energy theory community who engaged in the Snowmass process, participated in Theory Frontier activities, and provided feedback at various stages. We thank Masha Baryakhtar, Will Benoit, Ayres Freitas, Elvira G\'amiz, Steve Giddings, Stefania Gori, Gary Horowitz, Tristan Hubsch, Joshua Isaacson, Simon Knapen, Andreas Kronfeld, Eric Lindner, Yannick Meurice, Gil Paz, Gilad Perez, Michael Peskin, Chris Quigg, Tania Robens, Jon Rosner, Natalia Toro, and Jure Zupan for comments and feedback on this report.

The authors of this report were supported in part by grants and contracts from the U.S. Department of Energy, Office of Science, Offices of High Energy Physics and Nuclear Physics, from the U.S. National Science Foundation, from the Simons Foundation, the US-Israeli Binational Science Foundation, and from funding agencies in Japan and in Europe. We thank the Kavli Institute for Theoretical Physics for their generous support in hosting the Theory Frontier conference.

\bibliographystyle{JHEP}
\bibliography{Theory.bib}
\end{document}

%% file: workshopsymbols.tex
\def\ie{{\it i.e.}}

\def\beq{\begin{equation}}
\def\eeq#1{\label{#1}\end{equation}}
\def\eeqn{\end{equation}}

\newenvironment{Eqnarray}%
   {\arraycolsep 0.14em\begin{eqnarray}}{\end{eqnarray}}
\def\beqa{\begin{Eqnarray}}
\def\eeqa#1{\label{#1}\end{Eqnarray}}
\def\eeqan{\end{Eqnarray}}

\let\bar=\overbar

\def\lsim{\mathrel{\raise.3ex\hbox{$<$\kern-.75em\lower1ex\hbox{$\sim$}}}}
\def\gsim{\mathrel{\raise.3ex\hbox{$>$\kern-.75em\lower1ex\hbox{$\sim$}}}}

\def\del{\partial}
\def\Dslash{\not{\hbox{\kern-4pt $D$}}}
\def\dslash{\not{\hbox{\kern-2pt $\del$}}}
\def\pslash{\not{\hbox{\kern-2pt $p$}}}
\def\ETmiss{\not{\hbox{\kern-4pt $E$}}_T}

\def\Dlr{\mathrel{\raise1.5ex\hbox{$\leftrightarrow$\kern-1em\lower1.5ex\hbox{$D$}}}}

\def\MSB{{\bar{M \kern -2pt S}}}
\def\msb{{\bar{\scriptsize M \kern -1pt S}}}

\def\drb{{\bar{\scriptsize D \kern -1pt R}}}



%% file: authors.tex
\author[1]{\textbf{Frontier Conveners:} N.~Craig}
\author[2]{C.~Cs\'aki}
\author[3,4]{A.~X.~El-Khadra}
\author[5]{\\\ \authorcr\textbf{Topical Group Conveners:} Z.~Bern}
\author[6]{R.~Boughezal}
\author[7]{S.~Catterall}
\author[8]{Z.~Davoudi}
\author[9]{A.~de~Gouv\^ea}
\author[3,4]{P.~Draper}
\author[10]{P.~J.~Fox}
\author[11]{D.~Green}
\author[12]{D.~Harlow}
\author[10]{R.~Harnik}
\author[13]{V.~Hubeny}
\author[14]{T.~Izubuchi}
\author[15]{S.~Kachru} 
\author[16]{G.~Kribs}
\author[17,18,19]{H.~Murayama}
\author[19]{Z.~Ligeti}
\author[20]{J.~Maldacena}
\author[21,22]{F.~Maltoni}
\author[23]{I.~Mocioiu} 
\author[24]{E.~T.~Neil} 
\author[25,26]{S.~Pastore} 
\author[27]{D.~Poland}
\author[28]{L.~Rastelli}
\author[29]{I.~Rothstein}
\author[30]{J.~Ruderman}
\author[18,19]{B.~Safdi}
\author[3,4]{J.~Shelton}
\author[31]{L.~Strigari}
\author[32]{S.~Su}
\author[12]{J.~Thaler}
\author[13]{J.~Trnka}
\author[33]{\\\ \authorcr\textbf{Liaisons:} K.~Babu}
\author[34]{Steven Gottlieb}
\author[35,36]{A.~Petrov}
\author[37]{L.~Reina}
\author[38]{F.~Tanedo}
\author[39]{D.~Walker}
\author[40]{L.-T.~Wang}

\affil[1]{Department of Physics, University of California Santa Barbara, Santa Barbara, CA 93106, USA}
\affil[2]{Department of Physics, Cornell University, Ithaca, NY 14850, USA}
\affil[3]{Department of Physics, University of Illinois, Urbana, IL 61801, USA}
\affil[4]{Illinois Center for Advanced Studies of the Universe, University of Illinois, Urbana, IL 61801, USA}
\affil[5]{Mani L.~Bhaumik Institute for Theoretical Physics, UCLA Dept.~of Physics \& Astronomy, Los Angeles, CA 90095, USA}
\affil[6]{HEP Division, Argonne National Laboratory, Argonne, Illinois 60439, USA}
\affil[7]{Department of Physics, Syracuse University, Syracuse, NY 13244, USA}
\affil[8]{Department of Physics \& Maryland Center for Fundamental Physics, University of Maryland, College Park, MD 20742, USA}
\affil[9]{Northwestern University, Department of Physics \& Astronomy,  Evanston, IL 60208, USA}
\affil[10]{Theoretical Physics Department, Fermilab, Batavia, IL 60510, USA}
\affil[11]{Department of Physics, University of California San Diego, La Jolla, CA 92093, USA}
\affil[12]{Center for Theoretical Physics, Massachusetts Institute of Technology, Cambridge, MA 02139, USA}
\affil[13]{Center for Quantum Mathematics \& Physics (QMAP), University of California, Davis CA, USA}
\affil[14]{Physics Department, Brookhaven National Laboratory, Upton, NY 11973, USA}
\affil[15]{Stanford Institute for Theoretical Physics, Stanford University, Stanford, CA 94305, USA}
\affil[16]{Institute for Fundamental Science \& Department of Physics, University of Oregon, Eugene, OR 97403,  USA}
\affil[17]{Kavli Institute for the Physics \& Mathematics of the Universe (WPI), University of Tokyo Institutes for Advanced Study, University of Tokyo, Kashiwa 277-8583, Japan}
\affil[18]{Department of Physics, University of California, Berkeley, CA 94720, USA}
\affil[19]{Ernest Orlando Lawrence Berkeley National Laboratory, Berkeley, CA 94720, USA}
\affil[20]{Institute for Advanced Study, Princeton, NJ 08540}
\affil[21]{Centre for Cosmology, Particle Physics and Phenomenology (CP3), Universit\'{e} Catholique de Louvain, B-1348 Louvain la Neuve, Belgium}
\affil[22]{Dipartimento di Fisica e Astronomia, Universit\`{a} di Bologna and INFN, Sezione di Bologna, 40126 Bologna, Italy}
\affil[23]{Department of Physics, Pennsylvania State University, University Park, PA 16802, USA}
\affil[24]{Department of Physics, University of Colorado, Boulder, CO 80309, USA}
\affil[25]{Department of Physics, Washington University in Saint Louis, Saint Louis, MO 63130, USA}
\affil[26]{McDonnell Center for the Space Sciences at Washington University in St. Louis, MO 63130, USA}
\affil[27]{Department of Physics, Yale University, New Haven, CT 06520, USA}
\affil[28]{C. N. Yang Institute for Theoretical Physics, Stony Brook University, Stony Brook, NY
11794-3840, USA}
\affil[29]{Department of Physics, Carnegie Mellon University, Pittsburgh, PA 15213, USA}
\affil[30]{Center for Cosmology and Particle Physics, Department of Physics, New York University, New York, NY 10003, USA}
\affil[31]{Department of Physics \& Astronomy, Mitchell Institute for Fundamental Physics \& Astronomy, Texas A\&M University, College Station, TX 77843, USA}
\affil[32]{Department of Physics, University of Arizona, Tucson, AZ 85721, USA}
\affil[33]{Department of Physics, Oaklahoma State University, Stillwater, OK 74078, USA}
\affil[34]{Department of Physics, Indiana University, Bloomington, IN 47405, USA}
\affil[35]{Department of Physics, Wayne State University, Detroit, MI 48201, USA}
\affil[36]{Department of Physics \& Astronomy, University of South Carolina, Columbia, SC 29208, USA}
\affil[37]{Department of Physics, Florida State University, Tallahassee, FL 32306, USA}
\affil[38]{Department of Physics \& Astronomy, University of California Riverside, Riverside, CA 92521, USA}
\affil[39]{Department of Physics \& Astronomy, Dartmouth College, Hanover, NH 03755, USA}
\affil[40]{Department of Physics, University of Chicago, Chicago, IL 60637, USA}